\documentclass[a4paper,12pt]{article}
\usepackage{epsfig}
\usepackage[dvips,usenames]{color}
\usepackage{axodraw}
\usepackage{graphicx}

\newlength{\dinwidth}
\newlength{\dinmargin}
\date{}
\begin{document}
\title{Revisiting Charmless Two-Body B Decays involving $\eta^{\prime}$
and $\eta$
}
\author{Mao-Zhi Yang\\
\small{
Theory Division, Institute of High Energy Physics, Chinese Academy of Sciences,}\\
\small{P.O. Box 918(4), Beijing 100039, China}\\
\small{Physics Department, Hiroshima University, Higashi-Hiroshima,
Hiroshima 739-8526, Japan}\\
Ya-Dong Yang\\
\small{Physics Department, Technion-Israel Institute of Technology, Haifa 32000,
 Israel}\\
\small{Physics Department, Ochanomizu University, 2-1-1 Otsuka, Bunkyo-ku,
Tokyo 112-8610, Japan}\\
}
\maketitle
 \begin{picture}(0,0)
       \put(335,365){HUPD-0011}
       \put(335,385){OCHA-PP-166}
       \put(335,405){hep-ph/0012208}
      \end{picture}
\vspace{-10mm}
\begin{abstract}
We have studied charmless two-body B decays involving  $\eta$ and $\eta'$ in the
 framework of QCD improved factorization appraoch. The spectator hard scattering
 mechanism for $\eta'$ production have been re-examined and extended, which
 contributions
are incorporated consistently into the factorizable leading contributions.
It is found that the conventional mechanism would give
 $Br(B\to \eta' K)\sim 30\times 10^{-6}$
which agrees with predictions based on naive factorization approaches.
With the incorporation of spectator hard scattering mechanism  contributions,
theoretical predictions  are improved much and the bulk of $Br(B\to \eta' K)$
are  accommodated in the reasonable parameter space. We  have also presented
calculations of $g^{\ast}g^{\ast}-\eta^{(\prime)}$ transition form factors
within the standard hard scattering approach. It is  shown that  the new
contributions in the modes such as $B\to \eta' +vector$ and
$B\to \eta+ \pi, \rho, K^{(\ast)}$ are small. Direct CP-violation in those
 decay modes are predicted. It is shown that
the prospects for measuring direct CP-violation effects in $B^{\pm} \to$
$\eta' K^{\pm}$, $\eta' \pi^{\pm}$, $\eta \pi^{\pm}$,
 and $\eta K^{\pm}$ are  promising at B factories in the near future.

\end{abstract}

\newpage

\section{Introduction}

The first evidence of strong penguin was observed by CLEO\cite{cleo97} in 1997
with the announcement of
\begin{eqnarray}
Br(B^{-}\to \eta' K^{-})=(6.5^{+1.5}_{-1.4}\pm 0.9 )\times 10^{-5},\nonumber \\
Br(B^0 \to \eta' K^0 )= (4.7^{+2.7}_{-2.0}\pm0.9)\times 10^{-5},
\end{eqnarray}
which are $2\sim 4$ times larger than any  theoretical predictions existed
at that time. Driven by the unexpected large data, these decays modes
have been investigated extensively \cite{zhint,ggko,kagan,duyang}. As a result of the
investigations in past years, the contribution of the conventional mechanism
estimated by using
the naive factorization and effective Hamiltonian for $B$ decays would
account for 1/4$\sim$ 1/2 of the data. Some new mechanisms
are proposed to explain the unexpected large rates of B decays
to $K\eta'$. Namely

A)large intrinsic charm content of $\eta'$\cite{zhint}
through the chain $b\to c\bar{c}s \to \eta' s $ or through
$b\to c\bar{c}s \to s g^{\ast}g^{\ast} \to \eta' s $\cite{ggko};

B) strong penguin
$b\to s g$\cite{kagan} enhanced by new physics;

C) spectator hard-scattering mechanism through
$g^{\star} g^{\star}\to \eta'$\cite{duyang, ahmady}.

For type A mechanism, its magnitude is characterized by the parameter
$f^{c}_{\eta'}$ defined by the matrix
$<\eta'|\bar{c}\gamma_{\mu}\gamma_{5}c|0>=-if^{c}_{\eta'}p_{\mu}$.
To account for the large branching ratio of $B\to\eta' K$, $f^{c}_{\eta'}$
would be as large as $(50\sim 180)MeV$\cite{zhint}. However, the later
analyses have shown $f^{c}_{\eta'}$  as small as $few$ MeV
\cite{aligreub,FKS,sen}. It is also realized that strength of the process
 $b\to c\bar{c}s \to s g^{\ast}g^{\ast} \to \eta' s $ is very small\cite{pko}.
Generally compared with uncertainties in form factors and light quark masses
 in the estimations of nonleptonic
B decays,  the contribution of type A mechanism for $\eta'$ exclusive
production is probably safe to be neglected. For type B mechanism,
it would be very interesting to find signals of  new physics beyond
the standard model(SM) in these decays if  the SM
is indeed incapable of accommodating experimental data.
In this paper, we study those processes in the SM and investigate
the possibility whether
the SM can accommodate the present experimental data with new
approach for the hadronic dynamics of B decays.

For type C mechanism, it may be  promising.
Unfortunately it depends on some unknown quantities: the transition
form factor of $\eta'-g^{\ast}g^{\ast}$ and the light cone distribution
amplitudes(DA) of the mesons in the process. The prediction in
Ref.\cite{duyang} should be improved and incorporated with the
predictions of  the basic mechanisms consistently.

While the data of these decays reported in 1997\cite{cleo97}
is still puzzlingly large for theorists,
 robust experimental
investigations are in progress. Recently, using  the full CLEO II/II.V
data,
 CLEO Collaborations\cite{cleo2}
have improved their previous measurements of $B\to \eta' K $
\begin{eqnarray}
{\cal B}(B^+ \to \eta' K^{+})&=&(80^{+10}_{-9}\pm 7)\times 10^{-6},\nonumber \\
{\cal B}(B^0 \to \eta' K^{0})&=&(89^{+18}_{-16}\pm 9)\times 10^{-6},
\end{eqnarray}
with observations of two new decay modes
\begin{eqnarray}
{\cal B}(B^+ \to \eta K^{\ast+})&=&(26.4^{+9.6}_{-8.2}\pm 3.3)\times 10^{-6},
\nonumber\\
{\cal B}(B^0 \to \eta K^{\ast0})&=&(13^{+5.5}_{-4.6}\pm 1.6)\times 10^{-6},
\end{eqnarray}
and upper limits for other 12 decay modes involving $\eta$ or $\eta'$.

Theoretically, the importance and generality of the pioneer works of
Polizer and Wise\cite{wise} and factorization arguments of Bjorken\cite{bj}
are renewed by Beneke, Buchalla, Neubert and Sachrajda with the formation of
 ``QCD factorization" for B hadronic decays\cite{BBNS, BBNS2}. The factorization
formula incorporates elements of the naive factorization approach (as the leading
contribution) and the hard-scattering approach( as subleading
corrections), which allows us to calculate systematically radiative (subleading
nonfactorizable) corrections  to naive factorization for B exclusive nonleptonic
 decays.
An important product of the formula is that the strong final-state interaction
phases are calculable from the first principle which arise from the hard-scattering
kernel and hence process dependent. The strong phases are very important for
studying CP violation in B decays.
Detailed proofs and arguments could be found
in\cite{BBNS2}. Here we recall briefly the essence of the QCD factorization
formula as follows.

The amplitude of $B$ decays to two light mesons, say $M_1$ and $M_2$,
is obtained through the hadronic matrix element
$\langle M_{1}(p_{1}) M_{2}(p_{2}){\mid}{\cal O}_i {\mid}B(p)\rangle$,
here $M_{1}$ denotes the final meson that picks up the light spectator
quark in the $B$ meson, and $M_{2}$ is the other meson which is composed
of the quarks produced from the weak decay point of $b$ quark. Since the
quark pair, forming $M_2$, is ejected from the decay point of $b$ quark
carrying the large energy of order of $m_b$, soft gluons with the
momentum of order of $\Lambda_{QCD}$ decouple from it at leading order
of $\Lambda_{QCD}/m_b$ in the heavy quark limit. As a consequence any
interaction between the quarks of $M_2$ and the quarks out of $M_2$ is
hard at leading power in the heavy quark expansion. On the other hand,
the light spectator quark carries the momentum of the order of $\Lambda_{QCD}$,
and is softly transferred into $M_1$ unless it undergoes a hard interaction.
Any soft interaction between the spectator quark and other constituents in
$B$ and $M_1$ can be absorbed into the transition formfactor of $B\to M_1$ which
could be extracted from semileptonic decays $B\to M_1 l \bar{\nu}$.
The non-factorizable contribution to $B\to M_1 M_2$ can be calculated through
the diagrams in Fig.1, which turns out to be subleading oder corrections to
factorizable amplitudes.

In this paper we study $B\to \eta^{(\prime)} M (M=\pi,K^{(\ast)},\rho)$
decays within the framework of  QCD  factorization approach \cite{BBNS,BBNS2}.
We compare our numerical results with
the experimental data presented by CLEO collaboration \cite{cleo2}.
We find that the conventional mechanism contributions to
$B^{+}\to \eta' K^{+}$ and
$B^0 \to \eta' K^{0}$ are about $27\times 10^{-6}$ and $28\times 10^{-6}$
respectively, which agree with many theoretical expectations based on naive
factorizations. To explain the experimental data, contributions from new
mechanisms with the strength  as large as $40\%\sim 50\%$ of the strength of the
conventional mechanisms are found. Incorporating the contribution of spectator
hard-scattering mechanism(SHSM) to these decays, the experimental data could be
well accommodated in the SM. SHSM is found to be important
for $B\to\eta^{\prime}K$ but $not$  for $B\to\eta^{\prime}K^{\ast}$.

Our predictions agree with the data of  measured branching ratios
or lie   below their upper limits of other decay modes.
We also give our predictions of direct CP asymmetries
in these decay modes. Direct CP violations in the four
observed B decay modes $B^+ \to K^{+\ast}\eta$  and
$B^0 \to K^{0} \eta'(K^{0\ast} \eta)$ are found to be about a few percentages,
but it can reach $10\%$ for $B^{\pm} \to K^{\pm}\eta'$.
Considering its large branching ratio, we may expect that direct CP violation
effects in charmless B decays would be firstly observed
in $B^{\pm} \to K^{\pm}\eta'$.
Large direct CP violation asymmetries are predicted for
the decay modes $B^{\pm} \to \eta'\pi^{\pm}$, $\eta\pi^{\pm}$ and
$B^+ \to \eta K^{+}$. Prospects for
observing direct CP-violations in the these decay modes at B factories are
very promising.

This paper is organized as following. In Sec.2 we present notations and
calculations of the conventional mechanisms
 contributions to these decays. In Sec.3 we calculate $g^{\ast}g^{\ast}-\eta'$
transition
form factors using the standard hard scattering framework of Brodsky and
Lepage\cite{brod}. With the form factor, we re-examine the contribution of
the spectator hard scattering mechanism for $B\to \eta' K$\cite{duyang} and
generalize it to other 14  decay modes.
Section 4 contains our numerical results for the branching ratios
of two body charmless B decays involving $\eta$ and $\eta'$.
 Direct CP-violations in these decays are presented in Sec. 5.
Sec. 6 is the  summary of our investigations.

\section{Calculations of the conventional mechanisms }
The contribution of the conventional mechanisms are governed by the
 effective Hamiltonian for $B$ decays which is given by \cite{buras},
\begin{equation}
\label{heff}
{\cal H}_{eff}
=\frac{G_{F}}{\sqrt{2}} \left[ V_{ub} V_{uq}^*
\left(\sum_{i=1}^{2}
C_{i}O_{i}^{u}+
\sum_{i=3}^{10}
C_{i} \, O_i + C_g O_g \right)+
V_{cb} V_{cq}^*
\left(\sum_{i=1}^{2}
C_{i}O_{i}^{c}+\sum_{i=3}^{10}
C_{i}O_{i}+C_{g}O_g \right)
 \right],
\end{equation}
with the effective operators given by
\begin{equation}\begin{array}{llllll}
O_1^{u} & = &(\bar q_\alpha  u_\alpha)_{V-A}\cdot (\bar
u_\beta  b_\beta)_{V-A} ,
&O_2^{u} & = & ( \bar q_\alpha  u_\beta))_{V-A}\cdot( \bar
u_\beta  b_\alpha))_{V-A} , \\
O_1^{c} & = &(\bar q_\alpha  c_\alpha)_{V-A}\cdot (\bar
c_\beta  b_\beta))_{V-A} ,
&O_2^{c} & = & ( \bar q_\alpha  c_\beta)_{V-A}\cdot (\bar
c_\beta  b_\alpha )_{V-A}, \\
O_3 & = & (\bar q_\alpha  b_\alpha)_{V-A}\cdot \sum_{q'}(\bar
 q_\beta'  q_\beta' )_{V-A},   &
O_4 & = & (\bar q_\alpha  b_\beta))_{V-A}\cdot \sum_{q'}(\bar
q_\beta'  q_\alpha' )_{V-A}, \\
O_5 & = &( \bar q_\alpha  b_\alpha)_{V-A}\cdot \sum_{q'}(\bar
q_\beta'  q_\beta')_{V+A} ,   &
O_6 & = & (\bar q_\alpha  b_\beta)_{V-A}\cdot \sum_{q'}(\bar
q_\beta'  q_\alpha')_{V+A} , \\
O_7 & = & \frac{3}{2}(\bar q_\alpha  b_\alpha)_{V-A}\cdot
\sum_{q'}e_{q'}(\bar q_\beta'  q_\beta' )_{V+A},   &
O_8 & = & \frac{3}{2}(\bar q_\alpha  b_\beta)_{V-A}\cdot
\sum_{q'}e_{q'}(\bar q_\beta'  q_\alpha')_{V+A} , \\
O_9 & = & \frac{3}{2}(\bar q_\alpha  b_\alpha)_{V-A}\cdot
\sum_{q'}e_{q'}(\bar q_\beta'  q_\beta')_{V-A} ,   &
O_{10} & = & \frac{3}{2}(\bar q_\alpha  b_\beta)_{V-A}\cdot
\sum_{q'}e_{q'}(\bar q_\beta'  q_\alpha'\ )_{V-A},\\
O_g &=& (g_s/8\pi^{2}) \, m_b \, \bar{q}_{\alpha} \, \sigma^{\mu \nu}
      \, R  \, (\lambda^A_{\alpha \beta}/2) \,b_{\beta}
      \ G^A_{\mu \nu}~.
\label{operators}
\end{array}
\end{equation}
Here $q=d,s$  and $ q'\epsilon\{u,d,s,c,b\}$,
$\alpha$ and $\beta$ are the $SU(3)$ color indices and
$\lambda^A_{\alpha \beta}$, $A=1,...,8$ are the Gell-Mann matrices,
and $G^A_{\mu \nu}$ denotes the gluonic field strength tensor.
The Wilson coefficients evaluated at $\mu=m_b$ scale are\cite{buras}
\begin{equation}
\begin{array}{ll}
        C_1= 1.082, &
        C_2= -0.185,\\
        C_3=  0.014, &
        C_4= -0.035,\\
        C_5=  0.009, &
        C_6= -0.041,\\
        C_7= -0.002/137,&
        C_8=  0.054/137,\\
        C_9= -1.292/137,&
        C_{10}= 0.262/137,\\
        C_g=-0.143.&
\end{array}\label{ci}
\end{equation}

The non-factorizable contributions to $B\to M_1 M_2$ can be calculated through
the diagrams in Fig.1. The results of our calculations are summarized  compactly
by the following equations

\begin{eqnarray}
{\cal T}_{p}&=&\frac{G_F}{\sqrt{2}}\sum_{p=u,c}^{} V_{pq}^{*}V_{pb}
\biggl\{
    a_{1}^{p} (BM_{1},M_{2})(\bar{q}  u)_{V-A}\otimes (\bar{u}  b)_{V-A}\biggr.
+a_{2}^{p}(BM_{1},M_{2})(\bar{u}  u)_{V-A}\otimes (\bar{q}  b)_{V-A}
 \nonumber\\
 &+& a_{3}^{p}(BM_{1},M_{2})(\bar{q'}  q')_{V-A}\otimes (\bar{q}  b)_{V-A}
     + a_{4}^{p}(BM_{1},M_{2})(\bar{q}  q')_{V-A}\otimes (\bar{q'}  b)_{V-A}
 \nonumber\\
  &+&a_{5}^{p}(BM_{1},M_{2})(\bar{q'}  q')_{V+A}\otimes (\bar{q}  b)_{V-A}
    +a_{6}^{p}(BM_{1},M_{2})(-2)(\bar{q}q')_{S+P}\otimes (\bar{q'}b)_{S-P}
\nonumber \\
   &+& a_{7}^{p}(BM_{1},M_{2}) \frac{3}{2}e_{q'} (\bar{q'}  q')_{V+A}\otimes
             (\bar{q}  b)_{V-A}  \\
   &+& (-2)\left( a_{8}^{p}(BM_{1},M_{2}) \frac{3}{2}e_{q'}+a_{8a}(BM_{1},M_{2})\right)
(\bar{q}q')_{S+P}\otimes (\bar{q'}b)_{S-P}
\nonumber \\
 &+&a_{9}^{p}(BM_{1},M_{2})\frac{3}{2}e_{q'}(\bar{q'}  q')_{V-A}\otimes
            (\bar{q}  b)_{V-A} \nonumber \\
 &+& \biggl.\left( a_{10}^{p}(BM_{1},M_{2})
\frac{3}{2}e_{q'}+a_{10a}^{p}(BM_{1},M_{2})\right)
         (\bar{q}  q')_{V-A}\otimes (\bar{q'}  b)_{V-A}\biggr\},\nonumber
\label{tp}
\end{eqnarray}
where the symbol $\otimes$ denotes
$\langle M_1 M_2|j_2\otimes j_1|B \rangle \equiv \langle M_2|j_2|0\rangle
  \langle M_1|j_1|B\rangle $. $M_1$ represents the meson which picks up
 the spectator quark through this paper.
For $M_1$ a light $vector$ meson and
$M_2$ a light $pseudoscalar$ meson, the effective $a_{i}^p$'s which contain
next-to-leading order(NLO)  coefficients and ${\cal O}(\alpha_s )$
hard scattering  corrections are found to be
\begin{eqnarray}
 a_{1,2}^{c}(BV,P)&=&0, \quad a_{i}^{c}(BV,P)=a_{i}^{u}(BV,P)\equiv a_{i}(BV,P),
 i=3, 5, 7, 8, 9, 10, 8a, 10a. \nonumber \\
 a_{1}^{u}(BV,P)&=&C_{1}+\frac{C_{2}}{N}
 +\frac{\alpha_{s}}{4\pi}\frac{C_F}{N}C_{2}F_{P},\nonumber \\
 a_{2}^{u}(BV,P)&=&C_{2}+\frac{C_{1}}{N}
+\frac{\alpha_{s}}{4\pi}\frac{C_F}{N}C_{1}F_{P},\nonumber \\
 a_{3}(BV,P)&=&C_{3}+\frac{C_{4}}{N}
+\frac{\alpha_{s}}{4\pi}\frac{C_F}{N}C_{4}F_{P}, \nonumber \\
 a_{4}^{p}(BV,P)&=&C_{4}+\frac{C_{3}}{N}
+\frac{\alpha_{s}}{4\pi}\frac{C_F}{N}
\Biggl[ C_{3}(F_{P}+G_{P}(s_{q})+G_{P}(s_{b}))+C_{1}G_{P}(s_{p})
\Biggr. \nonumber \\
& &\left. +(C_{4}+C_{6})\sum_{f=u}^{b}G_{P}(s_{f})+C_{g}G_{P,g}\right],\nonumber\\
a_{5}(BV,P)&=&C_{5}+\frac{C_{6}}{N}+\frac{\alpha_{s}}{4\pi}\frac{C_F}{N}
C_{6}(-F_{P}-12),
\nonumber \\
a_{6}^{p}(BV,P)&=&C_{6}+\frac{C_{5}}{N}
+\frac{\alpha_{s}}{4\pi}\frac{C_F}{N}
\Biggl[
C_{1}G^{\prime}_{P}(s_{p})+C_{3}(G^{\prime}_{P}(s_{q})
+G^{\prime}_{P}(s_{b})) \Biggr. \nonumber \\
& &\left.
+(C_{4}+C_{6})\sum_{f=u}^{b}G^{\prime}_{P}(s_{f})
+C_{g}G^{\prime}_{P,g}\right], \nonumber \\
a_{7}(BV,P)&=&C_{7}+\frac{C_{8}}{N}+\frac{\alpha_{s}}
{4\pi}\frac{C_F}{N}C_{8}(-F_{P}-12),
\nonumber \\
a_{8}(BV,P)&=&C_{8}+\frac{C_{7}}{N}, \nonumber \\
a_{8a}(BV,P)&=&\frac{\alpha_{s}}{4\pi}\frac{C_F}{N}
\left[(C_{8}+C_{10})\sum_{f=u}^{b}\frac{3}{2}e_{f}G^{\prime}_{P}(s_{f})
+C_{9}\frac{3}{2}(e_{q}G^{\prime}_{P}(s_{q})+e_{b}G^{\prime}_{P}(s_{b}))
\right], \nonumber \\
 a_{9}(BV,P)&=&C_{9}+\frac{C_{10}}{N}+\frac{\alpha_{s}}{4\pi}
\frac{C_F}{N}C_{10}F_{P},
\nonumber \\
 a_{10}(BV,P)&=&C_{10}+\frac{C_{9}}{N}+\frac{\alpha_{s}}{4\pi}\frac{C_F}{N}C_{9}F_{P},
\nonumber \\
a_{10a}(BV,P)&=&\frac{\alpha_{s}}{4\pi}\frac{C_F}{N}
\left[(C_{8}+C_{10})\frac{3}{2}\sum_{f=u}^{b}e_{f}G_{P}(s_{f})
+C_{9}\frac{3}{2}(e_{q}G_{P}(s_{q})+e_{b}G_{P}(s_{b}))
\right],
\label{aeff1}
\end{eqnarray}
where $p=(u,c),\quad q=(d, s)$,  $q'=(u, d, s)$, and $f=(u, d, s, c, b)$.
$C_{F}=(N^2 -1)/(2N)$ and $N=3$
is the  number of colors.  The internal quark mass in the penguin diagrams enters as
$s_{f}=m_{f}^2/m_{b}^{2}$. $\bar{x}=1-x$ and $\bar{u}=1-u$.

\begin{eqnarray}
F_{P}&=&-12\ln\frac{\mu}{m_b } -18 + f_{P}^{I} + f_{P}^{II},  \\
f_{P}^{I}&=&\int_{0}^{1}dx g(x)\phi_{P}(x), {\hskip 15mm}
g(x)= 3\frac{1-2x}{1-x}\ln x -3i\pi, \nonumber \\
f_{P}^{II}&=&\frac{4\pi^2 }{N}\frac{f_{V}f_B }{ A_{0}^{V}(0)
M_{B}^2 } \int_{0}^{1}dz\frac{\phi_{B}(z)}{z}
\int_{0}^{1}dx\frac{\phi_{V}(x)}{x}
 \int_{0}^{1}dy\frac{\phi_{P}(y)}{y}, \\
G_{P, g} &=& -\int_{0}^{1}dx \frac{2}{\bar{x}}\phi_{P}(x), \\
G_{P}(s_{q}) &=&  \frac{2}{3} - \frac{4}{3}\ln\frac{\mu}{m_b} +
4\int_{0}^{1}dx \phi_{P}(x) \int_{0}^{1}du\quad u\bar{u}\ln
\left[s_{q} -u\bar{u}\bar{x} -i\epsilon \right],\\
G^{\prime}_{P, g} &=& -\int_{0}^{1}dx \frac{3}{2}\phi^{0}_{P}(x)=-\frac{3}{2}, \\
G^{\prime}_{P}(s_{q}) &=&  \frac{1}{3} - \ln\frac{\mu}{m_b} +
3\int_{0}^{1}dx \phi^{0}_{P}(x) \int_{0}^{1}du\quad u\bar{u}\ln
\left[s_{q} -u\bar{u}\bar{x} -i\epsilon \right],
\label{func1}
\end{eqnarray}

For $M_1$ is $pseudoscalar$ and $M_2$ is $vector$, the co-efficents are
\begin{eqnarray}
 a_{1,2}^{c}(BP, V )&=&0, \quad a_{i}^{c}(BP, V )=a_{i}^{u}(BP, V )
\equiv a_{i}(BP, V ),
i=3, 5, 6, 7, 8, 9, 10,  10a. \nonumber \\
 a_{1}^{u}(BP, V )&=&C_{1}+\frac{C_{2}}{N}
+\frac{\alpha_{s}}{4\pi}\frac{C_F}{N}C_{2}F_{V},\nonumber \\
 a_{2}^{u}(BP, V )&=&C_{2}+\frac{C_{1}}{N}
+\frac{\alpha_{s}}{4\pi}\frac{C_F}{N}C_{1}F_{V},\nonumber \\
 a_{3}(BP, V )&=&C_{3}+\frac{C_{4}}{N}
+\frac{\alpha_{s}}{4\pi}\frac{C_F}{N}C_{4}F_{V}, \nonumber \\
 a_{4}^{p}(BP, V )&=&C_{4}+\frac{C_{3}}{N}
+\frac{\alpha_{s}}{4\pi}\frac{C_F}{N}
\Biggl[
C_{3}(F_{V}+G_{V}(s_{q})+G_{V}(s_{b}))+C_{1}G_{V}(s_{p})
\Biggr. \nonumber \\
& &\Biggl. +(C_{4}+C_{6})\sum_{f=u}^{b}G_{V}(s_{f})+C_{g}G_{V,g}
    \Biggr],\nonumber\\
a_{5}(BP, V )&=&C_{5}+\frac{C_{6}}{N}
+\frac{\alpha_{s}}{4\pi}\frac{C_F}{N}C_{6}(-F_{V}-12),
\nonumber \\
a_{6}(BP, V )&=&C_{6}+\frac{C_{5}}{N}, \nonumber \\
a_{7}(BP, V )&=&C_{7}+\frac{C_{8}}{N}
+\frac{\alpha_{s}}{4\pi}\frac{C_F}{N}C_{8}(-F_{V}-12),
\nonumber \\
a_{8}(BP, V )&=&C_{8}+\frac{C_{7}}{N}, \nonumber \\
a_{9}(BP, V )&=&C_{9}+\frac{C_{10}}{N}
+\frac{\alpha_{s}}{4\pi}\frac{C_F}{N}C_{10}F_{V},
\nonumber \\
 a_{10}(BP, V )&=&C_{10}+\frac{C_{9}}{N}
+\frac{\alpha_{s}}{4\pi}\frac{C_F}{N}C_{9}F_{V},
\nonumber \\
a_{10a}(BP, V )&=&\frac{\alpha_{s}}{4\pi}\frac{C_F}{N}
\left[(C_{8}+C_{10})\frac{3}{2}\sum_{f=u}^{b}e_{f}G_{V}(s_{f})
+C_{9}\frac{3}{2}(e_{q}G_{V}(s_{q})+e_{b}G_{V}(s_{b}))
\right],
\label{aeff2}
\end{eqnarray}
where
\begin{eqnarray}
F_{V}&=&-12\ln\frac{\mu}{m_b } -18 + f_{V}^{I} + f_{V}^{II},  \\
f_{V}^{I}&=&\int_{0}^{1}dx g(x)\phi_{V}(x), {\hskip 15mm}
g(x)= 3\frac{1-2x}{1-x}\ln x -3i\pi, \nonumber \\
f_{V}^{II}&=&\frac{4\pi^2 }{N}\frac{f_{P}f_B }{ f_{+}^{B\to P}(0)
M_{B}^2 } \int_{0}^{1}dz\frac{\phi_{B}(z)}{z}
\int_{0}^{1}dx\frac{\phi_{P}(x)}{x}
 \int_{0}^{1}dy\frac{\phi_{V}(y)}{y}, \\
G_{V, g} &=& -\int_{0}^{1}dx \frac{2}{\bar{x}}\phi_{V}(x), \\
G_{V}(s_{q}) &=&  \frac{2}{3} - \frac{4}{3}\ln\frac{\mu}{m_b} +
4\int_{0}^{1}dx \phi_{V}(x) \int_{0}^{1}du\quad u\bar{u}\ln
\left[s_{q} -u\bar{u}\bar{x} -i\epsilon \right].
\label{func2}
\end{eqnarray}
For both $M_1$ and $M_2$ are $pseudoscalars$ , the co-efficents are
\begin{eqnarray}
 a_{1,2}^{c}(BM_{1}, M_2 )&=&0, \quad a_{i}^{c}(BM_{1}, M_2 )=a_{i}^{u}(BM_{1}, M_2 ),
i=3, 5, 7, 8, 9, 10, 8a, 10a. \nonumber \\
 a_{1}^{u}(BM_{1}, M_2 )&=&C_{1}+\frac{C_{2}}{N}
+\frac{\alpha_{s}}{4\pi}\frac{C_F}{N}C_{2}F_{M_{2}},\nonumber \\
 a_{2}^{u}(BM_{1}, M_2 )&=&C_{2}+\frac{C_{1}}{N}
+\frac{\alpha_{s}}{4\pi}\frac{C_F}{N}C_{1}F_{M_{2}},\nonumber \\
 a_{3}(BM_{1}, M_2 )&=&C_{3}+\frac{C_{4}}{N}
+\frac{\alpha_{s}}{4\pi}\frac{C_F}{N}C_{4}F_{M_{2}}, \nonumber \\
 a_{4}^{p}(BM_{1}, M_2 )&=&C_{4}+\frac{C_{3}}{N}
+\frac{\alpha_{s}}{4\pi}\frac{C_F}{N}
\Biggl[
C_{3}(F_{M_{2}}+G_{M_{2}}(s_{q})+G_{M_{2}}(s_{b}))+C_{1}G_{M_{2}}(s_{p})
\Biggr. \nonumber \\
& &\Biggl. +(C_{4}+C_{6})\sum_{f=u}^{b}G_{M_{2}}(s_{f})+C_{g}G_{M_{2},g}
   \Biggr],\nonumber\\
a_{5}(BM_{1}, M_2 )&=&C_{5}+\frac{C_{6}}{N}
+\frac{\alpha_{s}}{4\pi}\frac{C_F}{N}C_{6}(-F_{M_{2}}-12),
\nonumber \\
a_{6}^{p}(BM_{1}, M_2 )&=&C_{6}+\frac{C_{5}}{N}
+\frac{\alpha_{s}}{4\pi}\frac{C_F}{N}
\Biggl[
C_{1}G^{\prime}_{P}(s_{p})+C_{3}(G^{\prime}_{M_2}(s_{q})
+G^{\prime}_{M_2}(s_{b})) \Biggr. \nonumber \\
& &\left.
+(C_{4}+C_{6})\sum_{f=u}^{b}G^{\prime}_{M_2}(s_{f})
+C_{g}G^{\prime}_{M_{2},g}\right], \nonumber \\
a_{7}(BM_{1}, M_2 )&=&C_{7}+\frac{C_{8}}{N}
+\frac{\alpha_{s}}{4\pi}\frac{C_F}{N}C_{8}(-F_{M_{2}}-12),
\nonumber \\
a_{8}(BM_{1}, M_2 )&=&C_{8}+\frac{C_{7}}{N}, \nonumber \\
a_{8a}(BM_{1}, M_2 )&=&\frac{\alpha_{s}}{4\pi}\frac{C_F}{N}
\left[(C_{8}+C_{10})\sum_{f=u}^{b}\frac{3}{2}e_{f}G^{\prime}_{M_{2}}(s_{f})
+C_{9}\frac{3}{2}(e_{q}G^{\prime}_{M_{2}}(s_{q})+e_{b}G^{\prime}_{M_{2}}(s_{b}))
\right], \nonumber \\
a_{9}(BM_{1}, M_2 )&=&C_{9}+\frac{C_{10}}{N}
+\frac{\alpha_{s}}{4\pi}\frac{C_F}{N}C_{10}F_{M_{2}},
\nonumber \\
 a_{10}(BM_{1}, M_2 )&=&C_{10}+\frac{C_{9}}{N}
+\frac{\alpha_{s}}{4\pi}\frac{C_F}{N}C_{9}F_{M_{2}},
\nonumber \\
a_{10a}(BM_{1}, M_2 )&=&\frac{\alpha_{s}}{4\pi}\frac{C_F}{N}
\left[(C_{8}+C_{10})\frac{3}{2}\sum_{f=u}^{b}e_{f}G_{M_{2}}(s_{f})
+C_{9}\frac{3}{2}(e_{q}G_{M_{2}}(s_{q})+e_{b}G_{M_{2}}(s_{b}))
\right],
\label{aeff3}
\end{eqnarray}
where
\begin{eqnarray}
F_{M_{2}}&=&-12\ln\frac{\mu}{m_b } -18 + f_{M_{2}}^{I} + f_{M_{2}}^{II},
 \label{func0} \\
f_{M_{2}}^{I}&=&\int_{0}^{1}dx g(x)\phi_{M_{2}}(x), {\hskip 15mm}
g(x)= 3\frac{1-2x}{1-x}\ln x -3i\pi, \nonumber \\
f_{M_{2}}^{II}&=&\frac{4\pi^2 }{N}\frac{f_{M_1}f_B }{ f_{+}^{B\to M_{1}}(0)
M_{B}^2 } \int_{0}^{1}dz\frac{\phi_{B}(z)}{z}
\int_{0}^{1}dx\frac{\phi_{M_1}(x)}{x}
 \int_{0}^{1}dy\frac{\phi_{M_2}(y)}{y}, \\
G_{M_{2}, g} &=& -\int_{0}^{1}dx \frac{2}{\bar{x}}\phi_{M_2}(x), \\
G_{M_2}(s_{q}) &=&  \frac{2}{3} - \frac{4}{3}\ln\frac{\mu}{m_b} +
4\int_{0}^{1}dx \phi_{M_2}(x) \int_{0}^{1}du\quad u\bar{u}\ln
\left[s_{q} -u\bar{u}\bar{x} -i\epsilon \right],\\
G^{\prime}_{M_{2}, g} &=& -\int_{0}^{1}dx \frac{3}{2}\phi^{0}_{M_2}(x)=-\frac{3}{2}, \\
G^{\prime}_{M_2}(s_{q}) &=&  \frac{1}{3} - \ln\frac{\mu}{m_b} +
3\int_{0}^{1}dx \phi^{0}_{M_2}(x) \int_{0}^{1}du\quad u\bar{u}\ln
\left[s_{q} -u\bar{u}\bar{x} -i\epsilon \right],
\label{func3}
\end{eqnarray}

where $\phi_{P}(x)$ and $\phi^{0}_{P}(x)$ are the pseudoscalar meson's  twist-2
and twist-3 distribution amplitudes (DA) respectively. $\phi_{V}(x)=\phi_{V,\|}(x)$
is the leading twist DA for the longitudinally polarized vector meson states. We
have used the fact that light vector meson is longitudinally polarized in
$B\to PV$ decays. In the derivation of the effective coefficients
$a_i$'s we have used NDR scheme and assumption of
asymptotic DAs. The infrared divergences in $Fig.1.a-d$ are cancelled 
in their sum.

With the effective coefficients in Eqs.\ref{aeff1}, \ref{aeff2} and \ref{aeff3}
we can write down the decay amplitudes of the decay modes (we only list
four decay modes here which have been observed by CLEO. The other
decay modes are given in appendix A)
\begin{flushleft}
\begin{eqnarray}
{\cal M}( B^{+}\to\eta^{\prime}K^{+})&=&i\frac{G_{F}}{\sqrt{2}}
 f_{K}(m_{B}^{2}-m_{\eta'}^{2})F^{B\to\eta'}(m^2_{K})V_{cb}(1-\frac{\lambda^2}{2})
\biggl\{ R_{c}e^{i\gamma}
\Bigl[
a_{1}(X)+a_{4}^{u}(X)
\Bigr.
\biggr.\nonumber\\
&&+
R_{K} \bigl[ a_6^{u}(X)+a_{8}(X)+a_{8a} (X)
       \bigr] +a_{10}(X)+ a_{10a}(X)
 \Bigr]
\nonumber \\
&&+ a_{4}^{c}(X)+a_{10}(X) +a_{10a}(X)+
 R_{K}\bigl[ a_{6}^{c}(X)
+a_{8}(X) +a_{8a}(X)
 \bigr]
\biggl.
\biggr\}
\nonumber \\
&&+ i\frac{G_F}{\sqrt{2}}(m_{B}^{2}-m_{K}^2)
F^{B\to K}(m^2_{\eta'})V_{cb}(1-\frac{\lambda^2}{2})
\biggl\{
R_{c}e^{i\gamma}
 \Bigl[ a_{2}(Y)f^u_{\eta'}
\biggr.
\nonumber \\
&&+\bigl( a_{3}(Y)-a_{5}(Y)
   \bigr) (2f^u_{\eta'}+f^s_{\eta'})+
a_{4}^{u}(Y)f^s_{\eta'}
+R^s_{\eta'} \bigl( a_{6}^{u}(Y) \nonumber \\
&&-\frac{1}{2}a_{8}(Y)+
  a_{8a}(Y)\bigr) (f^s_{\eta'}-f^u_{\eta'})
+\frac{1}{2}\bigl( -a_{7}(Y)+a_{9}(Y) \bigr) (f^u_{\eta'}-f^s_{\eta'})
\nonumber \\
&&+\bigl( -\frac{1}{2}a_{10}(Y)+a_{10a}(Y)
    \bigr) f^s_{\eta'}
\Bigr]
+\bigl( a_{3}(Y)-a_{5}(Y) \bigr)(2f^u_{\eta'}+f^s_{\eta'})
+a_{4}^{c}(Y)f^s_{\eta'} \nonumber \\
&& +R^s_{\eta'}\bigl( a_{6}^{c}(Y)
-\frac{1}{2}a_{8}(Y)+a_{8a}(Y) \bigr)(f^s_{\eta'}-f^u_{\eta'})
\nonumber \\
&& \biggl.
+\frac{1}{2}\bigl( a_{9}(Y)+a_{7}(Y)\bigr)(f^u_{\eta'}-f^s_{\eta'})
-(\frac{1}{2}a_{10}(Y)-a_{10a}(Y))f^s_{\eta'}
\biggr\} .  \\
&& with\,\, X=B^-  \eta^{\prime},K^- \,\,\, and \,\,\, Y=B^- K^- , \eta^{\prime}.
\nonumber
\end{eqnarray}
\end{flushleft}

\begin{eqnarray}
{\cal M}( B^{0}\to\eta^{\prime}K^{0})
&=&
i\frac{G_F}{\sqrt{2}}(m_{B}^{2}-m_K^2)F^{B\to K}(m^2_{\eta'})
V_{cb}(1-\frac{\lambda^2}{2})
\biggl\{
 R_{c}e^{i\gamma}
         \left[ a_{2}(X)f^u_{\eta'}+a_{3}(X)(2f^u_{\eta'}+f^s_{\eta'})
         \right.
\biggr. \nonumber\\
&+& a_{4}^{u}(X)f^s_{\eta'}-a_{5}(X)(2f^u_{\eta'}+f^s_{\eta'})
+R^s_{\eta'}(a_{6}^{u}(X)-\frac{1}{2}a_{8}(X)+a_{8a}(X))(f^s_{\eta'}-f^u_{\eta'})
\nonumber \\
&+& \frac{1}{2}(-a_{7}(X)+a_{9}(X))(f^u_{\eta'}-f^s_{\eta'})+
(-\frac{1}{2}a_{10}(X)+a_{10a}(X))f^s_{\eta'}
\Bigl.
\Bigr]
\nonumber \\
&+&
\Bigl[ a_{3}(X)(2f^u_{\eta'}+f^s_{\eta'})+
a_{4}^{c}(X)f^s_{\eta'}-a_{5}(X)(2f^u_{\eta'}+f^s_{\eta'})
\nonumber\\
&+&
R^{s}_{\eta'}(a_{6}^{c}(X)-\frac{1}{2}a_{8}(X)+a_{8a}(X))
(f^s_{\eta'}-f^u_{\eta'})
\nonumber \\
&+& \frac{1}{2}\bigl( a_{9}(X) -a_{7}(X)+\bigr)(f^s_{\eta'}-f^u_{\eta'})+
\bigl( -\frac{1}{2}a_{10}(X)+a_{10a}(X) \bigr) f^s_{\eta'}\Bigr]
\biggl.
\biggr\}\nonumber \\
&+&
i\frac{G_F}{\sqrt{2}}f_{K}(m_{B}^{2}-m_{\eta'}^{2})
F^{B\to\eta'}(m^2_{K})V_{cb}(1-\frac{\lambda^2}{2})
\biggl\{
 R_{c}e^{i\gamma}
   \biggr.
\nonumber \\
 &&
\left[
     a_{4}^{u}(Y)+R_{K}(a_{6}^{u}(Y)-\frac{1}{2}a_{8}(Y)+a_{8a}(Y))
-\frac{1}{2}a_{10}(Y)+a_{10a}(Y)\right]
\nonumber \\
&+&a_{4}^{c}(Y)+R_{K}(a_{6}^{c}(Y)-\frac{1}{2}a_{8}(Y)+a_{8a}(Y))
-\frac{1}{2}a_{10}(Y)+
a_{10a}(Y)
\biggl.
\biggr\},\\
&& with\,\, X=B^0 K^0 , \eta^{\prime}\,\,\, and \,\,\,
Y=B^0  \eta^{\prime},K^0
. \nonumber
\end{eqnarray}

\begin{eqnarray}
{\cal M}( B^{+}\to\eta K^{\ast+})
&=&\frac{G_F}{\sqrt{2}}A_0^{B\to K^{\ast}}(m^2_{\eta}) m_{B}^{2}
V_{cb}(1-\frac{\lambda^2}{2})
        \biggl\{
            R_{c}e^{i\gamma}\left[
            a_{2}(X)f^{u}_{\eta}+a_{3}(X)(2f^{u}_{\eta}+f^{s}_{\eta})
         \right. \biggr.
\nonumber \\
&+& a^u_{4}(X)f^{s}_{\eta}
          -a_{5}(X)(2f^{u}_{\eta}+f^{s}_{\eta})-R^s_{\eta}
(a^u_{6}(X)-\frac{1}{2}a_{8}(X)+a_{8a}(X))(f^{s}_{\eta}-f^{u}_{\eta})
\nonumber \\
&+&\frac{1}{2}(a_{9}(X)-a_{7}(X))(f^{u}_{\eta}-f^{s}_{\eta})
-(\frac{1}{2}a_{10}(X)+a_{10a}(X))f^{s}_{\eta}
 \Bigr]
   +a_{3}(X)(2f^{u}_{\eta}+f^{s}_{\eta})
\nonumber \\
&+&a_4^{c}(X)f^{s}_{\eta}-a_{5}(X)(2f^{u}_{\eta}+f^{s}_{\eta})-R^s_{\eta}(a^c_{6}(X)
-\frac{1}{2}a_{8}(X)+a_{8a}(X))
(f^{s}_{\eta}-f^{u}_{\eta})
\nonumber \\
&+&\frac{1}{2}(a_{9}(X)-a_{7}(X))(f^{u}_{\eta}-f^{s}_{\eta})
-(\frac{1}{2}a_{10}(X)+a_{10a}(X))f^{s}_{\eta}
\biggl.
\biggr\} \nonumber \\
&+&\frac{G_F}{\sqrt{2}}F^{B\to\eta} f_{K^{\ast}}m_{B}^{2}
V_{cb}(1-\frac{\lambda^2}{2})
         \biggl\{ R_{c}e^{i\gamma}
             \left[a_{1}(Y)+a^u_{4}(Y)
             \right.
         \biggr.
\nonumber \\
&+&
\biggl. \left.
a_{10}(Y)+a_{10a}(Y) \right]
+a_{4}^{c}(Y)+a_{10}(Y)+a_{10a}(Y)\biggr\}, \\
&& with\,\, X=B^- K^{\ast-} , \eta^{\prime}
 \,\,\,
and \,\,\, Y=B^-  \eta^{\prime},K^{\ast-}
. \nonumber
\end{eqnarray}

\begin{eqnarray}
{\cal M}( B^{0}\to\eta K^{\ast0})
&=&\frac{G_F}{\sqrt{2}}A_0^{B\to K^{\ast}}(m^2_{\eta}) m_{B}^{2}
V_{cb}(1-\frac{\lambda^2}{2})
        \biggl\{
            R_{c}e^{i\gamma}\left[
            a_{2}(X)f^{u}_{\eta}+a_{3}(X)(2f^{u}_{\eta}+f^{s}_{\eta})
         \right. \biggr.
\nonumber \\
&+& a^u_{4}(X)f^{s}_{\eta}
          -a_{5}(X)(2f^{u}_{\eta}+f^{s}_{\eta})-R^s_{\eta}
(a^u_{6}(X)-\frac{1}{2}a_{8}(X)+a_{8a}(X))(f^{s}_{\eta}-f^{u}_{\eta})
\nonumber \\
&+&\frac{1}{2}(a_{9}(X)-a_{7}(X))(f^{u}_{\eta}-f^{s}_{\eta})
-(\frac{1}{2}a_{10}(X)+a_{10a}(X))f^{s}_{\eta}
   \Bigl.
\Bigr]
   +a_{3}(X)(2f^{u}_{\eta}+f^{s}_{\eta})
\nonumber \\
&+&a^4_{c}(X)f^{s}_{\eta}-a_{5}(X)(2f^{u}_{\eta}+f^{s}_{\eta})-R^s_{\eta}(a^c_{6}(X)
-\frac{1}{2}a_{8}(X)+a_{8a}(X))
(f^{s}_{\eta}-f^{u}_{\eta})
\nonumber \\
&+&\frac{1}{2}(a_{9}(X)-a_{7}(X))(f^{u}_{\eta}-f^{s}_{\eta})
-(\frac{1}{2}a_{10}(X)+a_{10a}(X))f^{s}_{\eta}
\biggl.
\biggr\} \nonumber \\
&+&\frac{G_F}{\sqrt{2}}F^{B\to\eta} f_{K^{\ast}}m_{B}^{2}
V_{cb}(1-\frac{\lambda^2}{2})
         \biggl\{ R_{c}e^{i\gamma}
             \left[ a^u_{4}(Y)-\frac{1}{2}
                              a_{10}(Y)+a_{10a}(Y) \right]
         \biggr.
\nonumber \\
\biggl.
&+&a_{4}^{c}(Y)-\frac{1}{2}a_{10}(Y)+a_{10a}(Y)\biggr\},\\
&& with\,\, X=B^0 K^{\ast} , \eta^{\prime}
 \,\,\,
and \,\,\, Y=B^0  \eta^{\prime},K^{\ast0}
. \nonumber
\end{eqnarray}

Where $R_c =\mid \frac{V_{us}V^{\ast}_{ub}}{V_{cs}V^{\ast}_{cb}}\mid
=\frac{\lambda}{1-\lambda^{2}/2}|\frac{V^{\ast}_{ub}}{V_{cb}}|$.
$V_{cb}, V_{us}$ and $V_{cs}$ are chosen to be real and $\gamma$ is the phase of
$V^{*}_{ub}$. $\lambda=|V_{us}|=0.2196$. We will present inputs and numerical
results for those magnitudes in Sec.4.

\section{Re-examination  of  the spectator-hard-scattering mechanism}

 \subsection{Calculation of $g^{\ast}g^{\ast}-\eta^{(\prime)}$ transition
  form factor $F_{g^{\ast}g^{\ast}-\eta^{(\prime)}}(Q_{1}^{2},Q_{2}^2 )$}

Recently Ref.\cite{ymz} studied $g^{\ast}g-\eta^{(\prime)}$
transition form factor $F_{g^{\ast}g-\eta^{(\prime)}}(Q_{1}^{2},Q_{2}^2=0 )$.
The same method can be easily used to calculate
$F_{g^{\ast}g^{\ast}-\eta^{(\prime)}}(Q_{1}^{2},Q_{2}^2 )$, where both
of the two gluons are off-shell. To keep the completeness of this paper
we recapitulate the main points of this calculation here.

The $\eta^{\prime}$ meson is mainly a flavor singlet meson. Because of its
singlet structure the $\eta^{\prime}$ meson may have gluonic content
through the QCD anomaly. The contribution of gluonic wave function to
the transition form factor $F_{g^{\ast}g-\eta^{(\prime)}}(Q_{1}^{2},Q_{2}^2=0 )$
has been tested, which is very small \cite{ymz}.  By analysing the gluonic
wave function, when both of the two gluons are off-shell, it will give an
extra scale suppression. So for the case of
$F_{g^{\ast}g^{\ast}-\eta^{(\prime)}}(Q_{1}^{2},Q_{2}^2 )$, it may be safe to
neglect the contribution of gluonic wave function of $\eta^{\prime}$ meson.

We take the $q\bar{q}-s\bar{s}$ mixing scheme for the $\eta^{(\prime)}$
meson in this calculation. Here $q\bar{q}$ means the light quark pair $u\bar{u}$
and $d\bar{d}$ \cite{FKS}.  In this mixing scheme the parton Fock state
decomposition can be expressed as
\begin{eqnarray}
 \mid \eta\rangle  &=& cos\phi \mid \eta_q\rangle -sin\phi \mid \eta_s\rangle,
\nonumber \\
 \mid \eta'\rangle &=& sin\phi \mid \eta_q\rangle +cos\phi \mid \eta_s\rangle,
\end{eqnarray}
where $\phi$ is the mixing angle, and $ \mid \eta_q\rangle \sim
\frac{1}{\sqrt{2}}\mid u\bar{u}+d\bar{d} \rangle$,
$\mid \eta_s\rangle \sim \mid s\bar{s}\rangle$.

The diagrams contributing to the transition form factor
$F_{g^{\ast}g^{\ast}-\eta^{(\prime)}}(Q_{1}^{2},Q_{2}^2 )$  are shown in Fig.3.
By direct calculation of these two diagrams the
$g^{\ast}g^{\ast}-\eta^{(\prime)}$ transition form factors
can be parameterized as
\begin{equation}
<g^{\ast}_{a}g^{\ast}_{b}|\eta^{(\prime)}>=-4\pi\alpha_{s}\delta_{ab}
i \epsilon_{\mu\nu\alpha\beta}Q_{1}^{\alpha}Q_{2}^{\beta}
F_{g^{\ast}g^{\ast}-\eta^{(\prime)}}(Q_{1}^{2},Q_{2}^2).
\end{equation}
and $F_{g^{\ast}g^{\ast}-\eta^{(\prime)}}(Q_{1}^{2},Q_{2}^2)$ 
is found to be
\begin{equation}
F_{g^{\ast}g^{\ast}-\eta^{(\prime)}}(Q_{1}^{2},Q_{2}^2 )=
\frac{1}{2N}\sum_{q=u,d,s}^{} f^{q}_{\eta^{(\prime)}}\int_{0}^{1}dx
\frac{\phi_{\eta^{(\prime)}}(x)}{(1-x)Q_{1}^2 +xQ_{2}^{2}
-x(1-x)m^{2}_{\eta^{(\prime)}}+i\epsilon }
+(x\to 1-x ),
\end{equation}
where the variables $f^{q}_{\eta^{(\prime)}}$ can be related to the decay
constants of $ \mid \eta_q\rangle$ and $ \mid \eta_s\rangle$
\begin{eqnarray}
f^{u}_{\eta^{\prime}} &=& \frac{f_q}{\sqrt{2}}\sin\phi,~~~~~~
f^{s}_{\eta^{\prime}} = f_s\cos\phi, \nonumber \\
f^{u}_{\eta} &=& \frac{f_q}{\sqrt{2}}\cos\phi,~~~~~~
f^{s}_{\eta} = -f_s\sin\phi.
\end{eqnarray}
The decay constants $f_q$, $f_s$ and the mixing angle $\phi$ have
been constrained from the available experimental data,
$f_q=(1.07\pm 0.02)f_{\pi}$, $f_s=(1.34\pm 0.06)f_{\pi}$,
$\phi=39.3^0\pm 1.0^0$ \cite{FKS}.

To the accuracy of this paper, $\phi_{\eta^{(\prime)}}(x)$
is taken to be the leading twist DAs as \cite{grozin}
$\phi_{\eta^{(\prime)}}(x)=6x(1-x)$.

The transition form factor will play a pivotal role in estimations of
gluonic exclusive production of $\eta^{\prime}$ and $\eta$.

\subsection{The magnitude of spectator-hard scattering mechanism for
$B\to \eta' M $}
In this subsection, we would re-calculate  the magnitude of spectator hard
scattering mechanism(SHSM ) for $B\to \eta' K $ proposed in Ref.\cite{duyang}
and generalize it to $B\to \eta^{(\prime)} M$ where M is a light
pseudoscalar or vector  meson.

The SHSM is described by the Feynman diagrams in Fig.3 where b quark decays
to s(d) quark and a $hard$ gluon. Since the virtuality of the gluon is
much larger than the typical scale of QCD boundstate $\Lambda_{QCD}$ $i.e.,$
$1/Q_{1}^2 << 1/\Lambda_{QCD}^2 $, it
would fluctuate into a small size fast flying color-octet
$\bar{q}q$ pair well before
it flys out of the light cloud of B, so it would hard interact with the
spectator. More arguments for validity of using perturbative QCD in this
calculation could be found in Ref.\cite{duyang}.

For M  a light pseudoscalar, the amplitudes $M_1$ for Fig.3.a and
$M_2$ for Fig.3.b
are calculated to be
\begin{eqnarray}
M1 &=&\frac{G_F}{\sqrt{2}}(-V^{\ast}_{tq}V_{tb})C_{g}
m_{b}f_{B}f_{P}(2f^{u}_{\eta^{(\prime)}} +f^{s}_{\eta^{(\prime)}})
\alpha^{2}_s \frac{i}{N_{c}^3 }\times (m_B F_{tw2} -\mu_P F_{tw3} ),
\label{m1}
\\
F_{tw2}^{\eta^{(\prime)}}&=&-8\int^{1}_{0}dzdy\phi_{B}(z)\phi^{as}_{P}(y)\frac{1-z}{y-z}
\left(
\frac{(Q_{1}\cdot Q_{2})^2}{Q^2_1 (Q^2_2+i\epsilon)}-1
\right)   \nonumber \\
&&\int^1_0 dx \frac{\phi^{as}(x)}{\bar{x}Q_1^2 +xQ_2^2 -x\bar{x}
m^2_{\eta^{(\prime)}}+i\epsilon }.\\
F_{tw3}^{\eta^{(\prime)}}&=&-8\int^{1}_{0}dzdy\phi_{B}(z)\phi_{P}^{0}(y)\frac{1-y}{y-z}
\left(
\frac{(Q_{1}\cdot Q_{2})^2}{Q^2_1 (Q^2_2+i\epsilon) }-1
\right)   \nonumber \\
&&
\int^1_0 dx \frac{\phi^{as}(x)}{\bar{x}Q_1^2 +xQ_2^2 -x\bar{x}
m^2_{\eta^{(\prime)}}+i\epsilon },\\
M2&=&\frac{G_F}{\sqrt{2}}\sum_{f}^{} V^{\ast}_{fq}V_{fb}C_{1}^{f}
f_{B}f_{P}(2f^{u}_{\eta^{(\prime)}} +f^{s}_{\eta^{(\prime)}})
\alpha^{2}_s \frac{i}{N_{c}^3 }\times F_{PP}^{\eta^{(\prime)}}(s_f ),\\
F_{PP}^{\eta^{(\prime)}}(s_f )&=& 16 \int^{1}_{0}dzdy\phi_{B}(z)\phi_{P}^{as}(y)\frac{1}{y-z}
\left(Q_1^2 -
\frac{(Q_{1}\cdot Q_{2})^2}{Q^2_2}
\right)
\nonumber \\
&&
\int^1_0 du u\bar{u}
\left(-1+
2\ln\frac{\mu}{m_b}-\ln(s_{f}-u\bar{u}\frac{Q_1^2}{m_b^2 }-i\epsilon)
\right)   \nonumber \\
&&
\int^1_0 dx \frac{\phi^{as}_{\eta^{(\prime)}}(x)}{\bar{x}Q_1^2 +xQ_2^2 -x\bar{x}
m^2_{\eta^{(\prime)}}+i\epsilon }.
\end{eqnarray}
When M is a light vector meson, the amplitudes $M_3$ and $M_4$ for Fig.3.a and Fig.3.b
respectively are
\begin{eqnarray}
M3&=&\frac{G_F}{\sqrt{2}}(-V^{\ast}_{tq}V_{tb})C_{g}
m_{b}f_{B}f_{V}(2f^{u}_{\eta^{(\prime)}} +f^{s}_{\eta^{(\prime)}})
\alpha^{2}_s \frac{1}{N_{c}^3 }m_B \times F_{BPV1}^{\eta^{(\prime)} },\\
F_{BPV1}^{\eta^{(\prime)}}&=&- 8\int^{1}_{0}dzdy
\phi_{B}(z)\phi_{V}^{as}(y)\frac{1-z}{y-z}
\left(
\frac{(Q_{1}\cdot Q_{2})^2}{Q^2_1 (Q^2_2+i\epsilon)}-1
\right)   \nonumber \\
&&
\times\int^1_0 dx \frac{\phi^{as}(x)}{\bar{x}Q_1^2 +xQ_2^2 -x\bar{x}
m^2_{\eta^{(\prime)}}+i\epsilon }. \\
M4&=&\frac{G_F}{\sqrt{2}}\sum_{f}^{} V^{\ast}_{fq}V_{fb}C_{1}^{f}
f_{B}f_{V}(2f^{u}_{\eta^{(\prime)}} +f^{s}_{\eta^{(\prime)}})
\alpha^{2}_s \frac{1}{N_{c}^3 }\times F_{BPV2}^{\eta^{(\prime)}}(s_f ),\\
F_{BPV2}^{\eta^{(\prime)}}(s_f)&=&16 \int^{1}_{0}dz\int^{1}_{0}dy
\phi_{B}(z)\phi_{V}^{as}(y)\frac{1}{y-z}
\left(Q_1^2 -
\frac{(Q_{1}\cdot Q_{2})^2}{Q^2_2}
\right)
\nonumber \\
&&
\times\int^1_0 du u\bar{u}
\left(-1+
2\ln\frac{\mu}{m_b}-\ln(s_{f}-u\bar{u}\frac{Q_1^2}{m_b^2 }-i\epsilon)
\right)   \nonumber \\
&&
\times\int^1_0 dx \frac{\phi^{as}_{\eta^{(\prime)}}(x)}{\bar{x}Q_1^2 +xQ_2^2 -x\bar{x}
m^2_{\eta^{(\prime)}}+i\epsilon }.
\label{m4}
\end{eqnarray}
Where $q=d,s$ and $f=u,c$. $Q_1 =(1-z)P_B -(1-y)P_M$ and $Q_2 =zP_B -yP_M$.

It should be noted that the effective vertex $b\to s g^{\ast}$ calculated from
full theory in Ref.\cite{soni} could not be used here, otherwise the contribution
of top quark
penguin will be double counted when amplitudes of SHSM are added to these
in Eq.\ref{tp}. Other four fermion operators, say, $O_3$, $O_4$ and $O_6$ can
also be
inserted in Fig.3. However, dominate  contributions come from the insertion
of $O_1^{u,c}$.

With note of
$$
\frac{(Q_{1}\cdot Q_{2})^2 }{Q^2_1 Q^2_2 }-1 \simeq -\frac{1}{4z(1-z)},
\,\,\,\,\,\,
Q^2_{1}-\frac{(Q_{1}\cdot Q_{2})^2 }{Q^2_2 }\simeq m^2_{B}\frac{(y-z)}{4z}
$$ and the DAs to be given in next section, we can see that
$F^{\eta^{(\prime)}}_{tw2}$,
$F^{\eta^{(\prime)}}_{PP}(s_f )$, $F^{\eta^{(\prime)}}_{BPV1}$,
and   $F^{\eta^{(\prime)}}_{BPV2}(s_f )$
are free of IR divergence. Since the twist-3 DA $\phi^{0}_{P}(y)$ for
pseudoscalar meson does not fall at end point, integration of $1/(y-z)$
 over $y$ results in $\ln(z)$, but such large logarithm will be smeared by
DA of B meson through the integration $\int \phi_{B}(z)\ln z dz$, which are shown
by  analytic expressions in Appendix.B.
Generally the integrations are free of IR divergence.

\section{Numerical calculations and discussions of results}

In the numerical calculations we use \cite{pdg}
$$\tau(B^+)=1.65\times 10^{-12}s, ~~~\tau(B^0)=1.56\times 10^{-12}s,$$
$$ M_B = 5.2792 { GeV},~~~~ m_b =4.8GeV, ~~~~m_c =1.4 GeV,$$
$$ m_s=80MeV,~~f_B = 0.180GeV, ~~f_\pi = 0.133{ GeV},$$
$$f_K=0.158 GeV,~~f_{K^*}=0.214GeV, ~~f_{\rho}=0.21GeV.$$
For the chiral enhancement factors for the pseudoscalar mesons,
we take
$$
R_{\pi^{\pm}}=
R_{K^{\pm,0}}\simeq 1.2,
$$
which are consistent with the values used in \cite{BBNS,chengy,hsw}, and
$$
R^s_{\eta^{(\prime)}}=\frac{m^2_{\eta^{(\prime)}}}{m_s m_b},
~~~\mu_{P}=\frac{m_b R_{P}}{2}.
$$

We take the leading-twist DA $\phi(x)$ and the twist-3 DA $\phi^0(x)$ of
light pseudoscalar and vector mesons as the asymptotic form \cite{wave}
\begin{equation}
\phi_{P,V}(x) =  6 x (1-x), \quad \phi^{0}_{P}(x) =1.
\label{phi}
\end{equation}
For the $B$ meson, its DA is modeled as \cite{KLS,LUY},
\begin{equation}
\phi_B(x) = N_{B} x^2(1-x)^2 \mathrm{exp} \left
 [ -\frac{M_B^2\ x^2}{2 \omega_{B}^2} \right],
\label{bwav}
\end{equation}
with  $\omega_{B}=0.3$ GeV, and $N_{B}$ is the
normalization constant to make $\int_{0}^{1} dx \phi_{B}(x) =1$.
$\phi_{B}(x)$  is strongly peaked around $x=0.075$, which is
consistent with the observation of Heavy Quark Effective Theory that the
wave function should be peaked around $\Lambda_{QCD}/M_{B}$. We get
the object $\int_{0}^{1}dx \phi_{B}(x)/x=14.62$ which is very near to
the argument of  $\int_{0}^{1}dx \phi_{B}(x)/x\simeq m_{B}/\lambda_{B}=17.6$
in Ref.\cite{BBNS}.

We have used the unitarity of the CKM matrix
$V_{uq}^* V_{ub} + V_{cq}^* V_{cb} + V_{tq}^* V_{tb}=0$  to decompose the amplitudes
into terms containing $V_{uq}^* V_{ub}$ and  $V_{cq}^* V_{cb}$, and
\begin{equation}\begin{array}{ll}
|V_{us}|=\lambda=0.2196,& |V_{ub}/V_{cb}|=0.085\pm 0.02, \\
|V_{cs}|=1-\lambda^{2}/2& |V_{ud}|=1-\lambda^{2}/2.
\end{array}
\end{equation}
We would use the latest CLEO results for $|V_{cb}|\cite{cleovcb, cleoetac}$
\begin{equation}
|V_{cb}|=0.0464\pm0.0020(stat.)\pm0.0021(syst.)\pm0.0021(theor.),
\end{equation}
and  leave the CKM angle $\gamma$ as a free parameter.

For the form factors for $B\to \pi, K, K^{\ast} $ and $\rho$,
we take the results of light-cone
sum rule \cite{PBVB, ball}
\begin{equation}
F^{B\to \pi^{\pm}}(0)=0.3,~~F^{B\to K}(0)=0.36,~~
A_0^{B\to\rho^{\pm}}(0)=0.372, ~~A_0^{B\to K^*}(0)=0.470.
\end{equation}
Compared with these rather well studied form factors, the form factors
for $B\to \eta^{(\prime)}$ are poorly known, which has hindered theoretical
predictions for B decays involving $\eta^{(\prime)}$ very much for a long time.
Neglecting $\eta^{(\prime)}$ masses effects, we  argue the following
scaling relations for from factors at large recoil point in the heavy quark limit
$m_{b}\to\infty $
\begin{equation}
\frac{F^{B\to\eta}(0)}{F^{B\to\pi}(0)}\simeq\frac{f^{u}_{\eta}}{f_\pi},
~~~~~
\frac{F^{B\to\eta'}(0)}{F^{B\to\pi}(0)}\simeq\frac{f^{u}_{\eta'}}{f_\pi}.
\end{equation}
We get
\begin{equation}
 F^{B\to\eta}(0)=0.176,~~
 F^{B\to\eta'}(0)=0.142.
\end{equation}
which agree well with the values $F^{B\to\eta}(0)=0.181$ and
$ F^{B\to\eta'}(0)=0.148$ in Ref.\cite{chengy, hsw}.

\begin{table}[htbp]
\caption{ Branching ratios (in units of $10^{-6}$) for B charmless
decays involving $\eta^{(\prime)}$.
Experimental results are taken from \cite{cleo2}. Our results are made  for
four cases with $\gamma=55^{\circ}$ and two cases with $\gamma=85^{\circ}$.
 The labels CM A and CM B represent
estimations of the conventional mechanism contributions(CM) with inputs
$F^{B\eta^{(\prime)}}(0)=0.142(0.176)$(A)
and $F^{B\eta^{(\prime)}}(0)=0.185(0.229)$(B) respectively. Columns with
CM A(B)+SH
represent our results
with the incorporation of spectator hard scattering mechanism.}
\begin{center}
    \begin{tabular}{l|llll|ll|lc}
        \hline
            \hline
Decay&  & $\gamma=55^{\circ}$ & & & $\gamma=85^{\circ}$ & &CLEO{\cite{cleo2}}\\
\cline{2-7}
Modes  & CM A
&CM A+SH
&CM B
&CM B+SH
&CM B
&CM B+SH
& $B$ or $90\%$ UL
\\
\hline
$ B^{+}\to \eta' K^+  $
& 27.1
& 63.4
& 32.8
& 68.9
& 35.4
& 69.3
& $80^{+10}_{-9}\pm 7$       \\
\hline
$ B^{0}\to \eta' K^0  $
& 28.4
& 67.1
& 35.0
& 74.8
& 34.5
& 73.1
& $89^{+18}_{-16}\pm 9$       \\
\hline
$ B^{+}\to \eta K^{\ast+}  $
& 5.40
& 5.24
& 6.88
& 6.73
& 9.62
& 9.44
& $26.4^{+9.6}_{-8.2}\pm 3.3$       \\
\hline
$ B^{0}\to \eta K^{\ast0}  $
& 7.40
& 7.27
& 9.96
& 9.79
& 9.82
& 9.65
& $13.8^{+5.5}_{-4.6}\pm 1.6$       \\
\hline
$ B^{+}\to \eta' \pi^+  $
& 3.78
& 7.60
& 6.12
& 10.8
& 4.86
& 11.0
& $<$12       \\
\hline
$ B^{0}\to \eta' \pi^{0}  $
& 0.13
& 0.51
& 0.16
& 0.52
& 0.20
& 0.74
& $<$5.7       \\
\hline
$ B^{+}\to \eta' K^{\ast+}  $
& 5.87
& 6.40
& 5.02
& 5.47
& 3.92
& 4.21
& $<$35       \\
\hline
$ B^{0}\to \eta' K^{\ast0}  $
& 3.22
& 3.68
& 2.10
& 2.49
& 2.06
& 2.42
& $<$24       \\
\hline
$ B^{+}\to \eta' \rho^+  $
& 3.73
& 3.71
& 6.36
& 6.33
& 6.54
& 6.60
& $<$33       \\
\hline
$ B^{0}\to \eta' \rho^{0}  $
& 0.020
& 0.010
&0.023
& 0.012
& 0.033
& 0.020
& $<$12       \\
\hline
$ B^{+}\to \eta K^{+}  $
& 3.11
& 7.85
& 1.49
& 6.00
& 1.21
& 4.69
& $<$6.9       \\
\hline
$ B^{0}\to \eta K^{0}  $
&  1.64
&  8.89
&  0.28
&  8.88
&  0.28
&  7.87
& $<$9.3       \\
\hline
$ B^{+}\to \eta \pi^{+}  $
& 6.54
& 4.84
& 10.2
& 8.02
& 7.96
& 5.15
& $<$5.7       \\
\hline
$ B^{0}\to \eta \pi^{0}  $
& 0.36
& 0.41
& 0.42
& 0.46
& 0.53
& 0.60
& $<$2.9       \\
\hline
$ B^{+}\to \eta \rho^{+}  $
& 6.45
& 6.39
& 10.7
& 10.6
& 10.3
& 10.4
& $<$15       \\
\hline
$ B^{0}\to \eta \rho^{0}  $
& 0.02
& 0.017
& 0.023
& 0.001
& 0.033
& 0.015
& $<$10       \\
\hline
\hline
    \end{tabular}
\end{center}
\label{tab2}
\end{table}

Taking $\gamma=55^{\circ}$ as benchmark, we present our numerical
results in Table.1. As shown in the table,  for the case of
$F^{B\eta^{(\prime)}}(0)=0.176(0.142)$,
our estimations of the conventional mechanism contributions to
 $B^{+}\to K^+ \eta'$ and
$B^0 \to \eta' K^0 $ are about $(30\sim40)\times 10^{-6}$ which confirm
many theoretical estimations in the literature\cite{chengy,AKL}.
It implies that  significant new contributions to those decays are
needed to interpret  the CLEO data.
Before going further,
we note that the theoretical predictions for $B\to \eta K^{\ast}$
are also much smaller than their experimental results.
In contrast to
$B\to \eta' K$, new contributions to B decays $\eta K^{\ast}$
from both $b\to s(c{\bar c})\to s \eta$
and SHSM are negligible.
Because $\eta$ and $\eta'$ mesons are much heavier than $\pi$,
the recoiling of $\eta^{(\prime)}$ should be smaller than that of
$\pi$, so the scaling relation of Eq.49 might be broken to certain
extent.
Driven by the facts,
we pose $30\%$ enhancement to the form factor $F^{B\to\eta^{(\prime)} }(0)$.
From column 4 of Table.I, we can see the $30\%$ enhancement of
$F^{B\to\eta^{(\prime)}}(0)$ is  preferred by  experimental data.
However, conventional mechanism contributions to $B^{+}\to\eta' K^{+}$ and
$B^{0}\to\eta' K^{0}$  are just enhanced by $17\%$ and
still around 1/2 of their experimental center
values. With incorporation of SHSM contributions, theoretical results
are improved much, and the experimental data from CLEO could be accommodated,
although our results are still slightly lower than the center values.
It is also noted that BaBar collaborations\cite{BaBar} have reported their
preliminary results
\begin{equation}
{\cal B}(B^{+}\to \eta' K^{+} )=(62\pm 18\pm 8)\times 10^{-6},
\,\,\, {\cal B}(B^{0}\to \eta' K^{0} )<112\times 10^{-6}
\end{equation}
where the center value of ${\cal B}(B^{+}\to \eta' K^{+} )$ is smaller than
that of CLEO's. It might be safe to conclude  that the large branching ratio of
$B\to\eta'K$ could be understood in the SM.

Topologically, SHSM would contribute to the decays $B\to \eta' K^{\ast}$ the same
as it to
$B\to \eta' K$. However, SHSM amplitudes depend on the spin configurations
of $K$ and $K^{\ast}$. As shown in $Eqs.\ref{m1}-\ref{m4}$, compared with
the SHSM amplitudes
for $B\to K^{\ast}\eta'$, SHSM amplitudes for $B\to \eta' K$ are chirally enhanced
by $K$ twist-3 DA. So that the new contributions to $B\to \eta' K $ are
much larger than its to $B\to \eta' K^{\ast}$. In Fig. 4, we display our results
for the branching ratios of $B$ two-body charmless decays involving $\eta^{(\prime)}$
as functions of the weak angle $\gamma$. From  Fig.4, we can see that
SHSM  contributions to $B \to \eta'+P$ are much larger than its
contributions to $B\to \eta'+V$. It is easy to understand that
SHSM contributions to $B\to \eta + M$ are very small because of cancellations between
$(u\bar{u}+d\bar{d})$ and $s\bar{s}$ contents of $\eta$.
For the four measured decays $B^{+}\to (\eta' K^{+},~\eta K^{\ast})$ and
$B^{0}\to (\eta' K^{0},\eta K^{0\ast})$, our results generally agree with
CLEO data. In our calculations, we have used large form factor $F^{B\to\eta}(0)$ to
give large ${\cal B}(B^+ \to \eta K^{\ast+}) $ and ${\cal B}(B^0 \to \eta K^{\ast0})$,
meanwhile  we meet constraint from $B^{+}\to\eta\pi^{+}$ whose  upper limit
set by CLEO is ${\cal B}(B^+ \to\eta\pi^+ )<5.7\times 10^{-6}$.
Since  $B^{+}\to\eta\pi^{+}$ is dominated by
the tree amplitude with $if_{\pi}F^{B\to\eta}(m^2_{\pi})a_1$,
 ${\cal B}(B^{+}\to\eta\pi^{+})$
is very sensitive to the form factor $ F^{B\to\eta}(m^2_{\pi})$.  To meet the
large  branching ratio
 ${\cal B}(B^+ \to \eta K^{\ast+})=(26.4^{+9.6}_{-8.2}\pm3.3)\times10^{-6}$
and  comply with the small upper limit
${\cal B}(B^{+}\to\eta\pi^{+})<5.7\times 10^{-6}$, we need $\cos\gamma<0$ to
enhance
 ${\cal B}(B^+ \to \eta K^{\ast+})$ and suppress
${\cal B}(B^{+}\to\eta\pi^{+})$. This situation is very similar to that in
$B^{0}({\bar B}^{0}) \to \pi^{\pm}\pi^{\mp},~ K^{\pm}\pi^{\mp}$
and $B\to\pi\rho, \pi K^{\ast}$ decays as discussed in
Refs.\cite{chengy, he, desh, houyang,  gro, yang} recently.
It is worth to note that
most recent theoretical analyses of CLEO data  based on different
approaches  endorse, although not
definitely,   negative $\cos\gamma$ to some extent. However global CKM fit
has given the constraint $\gamma < 90^{\circ}$ at $99.6\%$ C.L.
\cite{ckmfit,ckm2000}.
With refined measurements at running B factories BELLE and BaBar to come very
soon, the following
B exclusive decay modes will give strong constraints on $\gamma$
\begin{eqnarray}
PP~~ modes:& B^0 \to \pi^{\pm}\pi^{\mp},~ B^0 \to K^{\pm}\pi^{\mp},~
          B^{\pm}\to K^{\pm}\pi^{0},~ B^{\pm} \to \eta' \pi^{\pm},~
          B^{\pm}\to \eta\pi^{\pm};\label{chanceA}\\
PV~~ modes:& B^0 \to \pi^{\pm}\rho^{\mp},~ B^{\pm} \to\pi^{-}\omega,~
           B^{\pm} \to \pi^{\pm}\rho^{0},~ B^{\pm}\to K^{\pm}\omega,~
           B^{\pm} \to \eta K^{\ast\pm}.
\label{chanceB}
\end{eqnarray}
Branching ratios for the above decay modes are of order of $10^{-6}\sim 10^{-5}$
which can be well studied at B factories to constrain  $\cos\gamma$ tightly.
If the disagreement of constraints on $\gamma$ between global fit and direct model
calculations becomes serious, it might imply the failure of the models employed
here and in Refs.\cite{chengy, he, desh, houyang,  gro, yang}.
Very probably theories for B hadronic decays will be tested and driven   by
the observations to be made at BaBar and Belle.

\section{Final states interactions and CP violation}
As shown by Beneke, Buchalla, Neubert and Sachrajda in Refs.\cite{BBNS, BBNS2},
the final states interactions in charmless B two-body decays are calculable
in the QCD improved factorization framework, which turn out to be nonfactorizable
corrections. The nonfactorizable corrections for the decays studied in this paper
are shown in  Eqs.\ref{aeff1}, \ref{aeff2},\ref{aeff3}. SHSM amplitudes are
generally nonfactorizable and always contribute large strong phase to certain
decay modes.
It is worth to note that some shortcomings in the ``generalized factorization"
are resolved in the framework employed in this paper. Nonfactorizable effects
are calculated in a rigorous way here instead of being parameterized by
effective color number. Since the hard scattering kernels are convoluted
with the  light cone DAs of the mesons, gluon virtuality $k^2 ={\bar x}m_b^2$
in the penguin diagram Fig.1.e has well defined meaning and leaves no
ambiguity as to the value of $k^2$, which has conventionally been treated as a free
phenomenological parameter in the estimations of the strong phase generated
through the Bander, Silverman and Soni(BSS) mechanism\cite{BSS}.
So that CP asymmetries are predicted
soundly in this paper.

The direct CP asymmetry parameter is defined as
\begin{equation}
A^{dir}_{CP}=\frac{|{\cal M}(B^-\to \bar{f})|^2-|{\cal M}(B^+\to f)|^2}
                      {|{\cal M}(B^+\to f)|^2+|{\cal M}(B^-\to \bar{f})|^2},
\label{dircp1}
\end{equation}
and
\begin{equation}
A^{dir}_{CP}=\frac{ |{\cal M}(\bar{B}^0\to \bar{f})|^2
-|{\cal M}(B^0\to f)|^2 }
{|{\cal M}(B^0\to f)|^2+|{\cal M}(\bar{B}^0\to \bar{f})|^2}.
\label{dircp2}
\end{equation}
For CP-violations in $B^{0}(\bar{B}^0 )\to \eta^{(\prime)}\pi^{0},
\eta^{(\prime)}\rho^{0}$,
$B^{0}-\bar{B}^0$ mixing effects should be considered. However, the branching
ratio for these decay modes are very small( below $10^{-6}$). We would not address
CP violations in those decay modes in this paper.

Our numerical results for the direct CP violations are shown in Fig.5 as functions
of
$\gamma$. For $\gamma\in[50^{\circ}, 90^{\circ}]$, the direct CP-violation
in $B^{\pm}\to \eta' K^{\pm}$ are found about $8\% \sim 10\% $. The recent
search for direct CP violation in  $B^{\pm}\to \eta' K^{\pm}$ made by
CLEO Collaborations\cite{cleocp}
has reported
\begin{equation}
A_{CP}^{dir}(B^{\pm}\to \eta' K^{\pm})\sim +0.03\pm 0.12
\end{equation}
which results in the $90\%~C.L$ interval [-0.17, 0.23]
for $A^{dir}(B^{\pm}\to \eta' K^{\pm})$. With more and better data to come soon,
the sensitivity of $A^{dir}(B^{\pm}\to \eta' K^{\pm})$ at B factories could
reach $\sim\pm 4\%$. The prospect of observing direct  CP violation in
$B^{\pm}\to \eta' K^{\pm}$  are expected
to be quite good.
The direct CP violations in $B^{0}\to\eta' K^{0},~\eta K^{\ast}$ and
$B^{+} \to \eta K^{\ast+} $ are also estimated to be about few
percent but smaller than $A^{dir}(B^{\pm}\to \eta' K^{\pm})$.
 Considering their experimental sensitivities and/or branching ratios,
prospects of observing direct CP violations in these decay modes are much
weaker than in $B^{\pm}\to \eta' K^{\pm}$.

Dighe, Gronau and Rosner\cite{dgr} have predicted large CP violations in
$B^{\pm}\to\eta\pi^{\pm}$ and  $B^{\pm}\to\eta'\pi^{\pm}$. More earlier
similar conclusion could be found in Ref.\cite{sehgal}.
For $\gamma\in[50^{\circ}, 90^{\circ}]$,
 We find
\begin{equation}
{\cal A}_{CP}^{dir}(B^{+}\to\eta'\pi^{+})\approx -50\%\sim -80\%,\,\,
{\cal A}_{CP}^{dir}(B^{+}\to\eta\pi^{+})\approx +15 \%\sim +30\%.
\end{equation}
     From Fig.5.5 and Fig.5.11, we can see that SHSM contributions
could enhance   ${\cal A}^{dir}(B^{+}\to\eta^{(\prime)}\pi^{+})$ very much.
 As shown in Fig.4.5, the branching ratio
${\cal B}^{dir}(B^{+}\to\eta'\pi^{+}) $ is predicted to be of order of
$10^{-6}$.

It should be  very promising to observe direct CP violation in
$B^{\pm}\to\eta'\pi^{\pm},\, \eta\pi^{\pm}$ in the near future.
We also predict large
${\cal A}^{dir}(B^{+}\to\eta K^{+})$. The decay modes get a large
strong phase through SHSM as shown in Fig.4.11. The strong interference
between tree and penguin amplitudes leads to
\begin{equation}
{\cal A}^{dir}(B^{+}\to\eta K^{+})\simeq +20\%\sim +50\%
\end{equation}
for $\gamma\in [50^{\circ}, 90^{\circ}]$.

 To summarize this section, we find large  direct CP violations in decays
 $B^{\pm}\to \eta'K^{\pm}$,
 $B^{\pm}\to \eta'\pi^{\pm}$, $\eta\pi^{\pm}$  and
$B^{\pm}\to \eta K^{\pm}$.

\section{Summary}
With the newly developed QCD improved factorization approach\cite{BBNS,BBNS2},
we have  studied hadronic  charmless two-body decays of $B_u$ and $B_d$ involving
$\eta$ or $\eta'$. Nonfactorization effects are calculated in terms of
order of ${\cal O}(\alpha_{s})$ corrections to the leading factorizatable
amplitudes. We find again that the conventional mechanism account for
about one half of the decay rates of $B \to \eta' K$. Significant
contributions are needed to solve the ``puzzle" of $unexpected$ large branching
ratios.

To clarify possible sources of new significant contributions to
${\cal B}(B \to \eta' K)$ up to date, we display the following facts.
Theoretically new contributions due to intrinsic charm content of
the $\eta'$ have been realized to be  small. Motivated by their
observation of large B decay rates to $\eta' K$\cite{cleo97, cleo2},
CLEO Collaborations have
 made searching for
B decays to $\eta_c K$\cite{cleoetac}. It is found that there is no
unexpected enhancement in the $\eta_c K $ rate. Since the $\eta_c K$ rate
should also be enhanced if the $\eta' K$ rate enhanced by the intrinsic charm
 content of the $\eta'$, the CLEO results indicate that the charm content of the
$\eta'$ may be not the explanation for the anomalous ${\cal B}(B\to \eta' K)$.
Alternatively one may turn to new physics. We have known that the experimental
observations of $B\to X_s \gamma$ and $B\to K^{\ast}\gamma$  agree with the SM
expectations,    which implies that
there is no $anomalous$  $large$ new physics
effects in B decays. Motivated by these considerations, we have re-examined
the SHSM. Starting from calculations of the transition form-factor
$g^{\ast}g^{\ast}-\eta'$, we have incorporated the new contributions to
that of conventional mechanism for B decays and successfully accommodated
the large rate of  B decays to $\eta' K$ in the SM. We have found that
the absolute strength of the amplitude of SHSM for  B decays to $\eta' K$
are about $1/3 \sim 1/2$ of the absolute strength of the conventional mechanism
which shows the conventional mechanism to be the dominant.
As shown in detail, the SHSM is  important for B decays to $\eta' K$
but not for B decays to $\eta' K^{\ast}$.

We have estimated the direct CP violations in the decays.
$A^{dir}(B^{+} \to \eta' K^{+})$ is found to be around $+8\%\sim +10\%$
for $\gamma\in[50^{\circ}, 90^{\circ}]$. Due to significant  contributions
from SHSM, we predict
$$
A^{dir}(B^{\pm}\to\eta'\pi^{\pm})\simeq 40\%\sim 70\%. \,\,
A^{dir}(B^{\pm}\to\eta\pi^{\pm})\simeq 20\%\sim 40\%. \,\,
$$
We also predict large direct CP violation in B decays to $\eta K^{\pm}$.
Prospects for observing direct CP violations in $B^{\pm}$ decays to
$\eta'K^{\pm}$, $\eta'\pi^{\pm}$, $\eta\pi^{\pm}$ and $\eta K^{\pm}$ are
quite promising at the ruining B factories BaBar and BELLE.

$Note~added $ After we finished this work, we note  the work\cite{ali3}
by Ali and Parkhomenko which is focused on $g^{\ast}g^{\ast}-\eta'$ transistion
 form factor. They find that the contribution gluonic content of $\eta'$
to the form factor $F_{g^{\ast}g^{\ast}\eta'}(Q^2_{1}, Q^2_2 )$ as large as
few ten percentages for small $Q^2$ and rather small for large $Q^2$.
 Our formfactor agree with theirs to leading
terms. Additionally, in the framwork\cite{BBNS, BBNS2}
employed here all diagrames in Fig.1 and Fig.3 are subleadling nonfactorizable
contributions, so the gluonic contributions  could  be neglected.


\section*{Acknowledgments}

We acknowledge the Grant-in-Aid for Scientific Research on Priority Areas
(Physics of CP violation with contract number 09246105 and 10140028).
We thank JSPS(Japan Society for the Promotion of Science) for support.
The work of Y.Y is partially supported by the US-Israel Binational Science
Foundation and the Israel Science Foundation. Y.Y thanks Prof. Eilam and 
Prof. Gronau for many valuable discussions

\newpage

\begin{center} {\bf Appendix A}  \end{center}

The decay amplitudes the conventional mechanism of some of the  decays in terms of
the effective coefficients $a_i$'s are presented as the followings,

\begin{scriptsize}
\begin{flushleft}
\begin{eqnarray}
{\cal M}( B^{+}\to\eta^{\prime}\pi^{+})&=&i \frac{G_{F}}{\sqrt {2}} M_{B}^{2}
         F^{B\to\pi}(m^{2}_{\eta^{(\prime)}}) \lambda V_{cb}
   \biggl\{
         R_{u} e^{i\gamma} \Bigl[
          a_{2}(B^{+}\pi^{+},\eta^{\prime})f^{u}_{\eta^{(\prime)}}
         +[a_{3}(B^{+}\pi^{+},\eta^{\prime})-a_{5}(B^{+}\pi^{+},\eta^{\prime})]
           (2f^{u}_{\eta^{(\prime)}}+f^{s}_{\eta^{(\prime)}})
    \Bigr.
    \biggr.  \nonumber \\
     && +a_{4}^{u}(B^{+}\pi^{+},\eta^{\prime})  f^{u}_{\eta^{(\prime)}}
     +\bigl[a_{6}^{u}(B^{+}\pi^{+},\eta^{\prime})
      -\frac{1}{2}a_{8}(B^{+}\pi^{+},\eta^{\prime})
      +a_{8a}(B^{+}\pi^{+},\eta^{\prime})
       \bigr] R^{d}_{\eta^{(\prime)}} f^{u}_{\eta^{(\prime)}} \nonumber \\
      &&+
       \frac{1}{2}\bigl[ a_{9}(B^{+}\pi^{+},\eta^{\prime})-a_{7}(B^{+}\pi^{+},\eta^{\prime})
                   \bigr]
       (f^{u}_{\eta^{(\prime)}}-f^{s}_{\eta^{(\prime)}})
       +
         [-\frac{1}{2}a_{10}(B^{+}\pi^{+},\eta^{\prime})
         +a_{10a}(B^{+}\pi^{+},\eta^{\prime})] f^{u}_{\eta^{(\prime)}}
         \Bigr] \nonumber \\
       &&  -\Bigl[
            \bigl[ a_{3}(B^{+}\pi^{+},\eta^{\prime})-a_{5}(B^{+}\pi^{+},\eta^{\prime})
            \bigr]
        (2f^{u}_{\eta^{(\prime)}}+f^{s}_{\eta^{(\prime)}})
         +a_{4}^{c}(B^{+}\pi^{+},\eta^{\prime})  f^{u}_{\eta^{(\prime)}}
         +\bigl[ a_{6}^{c}(B^{+}\pi^{+},\eta^{\prime})
          \bigr.  \Bigr.       \nonumber \\
      && -\frac{1}{2}a_{8}(B^{+}\pi^{+},\eta^{\prime})
           +a_{8a}(B^{+}\pi^{+},\eta^{\prime}) \bigr]
           R^{d}_{\eta^{(\prime)}} f^{u}_{\eta^{(\prime)}}+
         \bigl[ a_{9}(B^{+}\pi^{+},\eta^{\prime})-a_{7}(B^{+}\pi^{+},\eta^{\prime})
         \bigr]
          \frac{1}{2}(f^{u}_{\eta^{(\prime)}}-f^{s}_{\eta^{(\prime)}}) \nonumber \\
      &&\biggl.  +
         \bigl[-\frac{1}{2}a_{10}(B^{+}\pi^{+},\eta^{\prime})
           +a_{10a}(B^{+}\pi^{+},\eta^{\prime})
         \bigr] f^{u}_{\eta^{(\prime)}}
         \Bigr]
        \biggr\} \nonumber \\
   && +i \frac{G_{F}}{\sqrt {2}} (M_{B}^{2}-m_{\eta^{(\prime)}}^{2})
F^{B\to\eta^{(\prime)}}(0) f_{\pi} \lambda V_{cb}
      \biggl\{
            R_{u} e^{i\gamma}
         \Bigl[ a_{1}(B^{+}\eta^{\prime},\pi^{+}) +
            a_{4}^{u}(B^{+}\eta^{\prime},\pi^{+})
          +\bigl[ a_{6}^{u}(B^{+}\eta^{\prime},\pi^{+}) \bigr. \biggr. \nonumber \\
   && \Bigl.
       +a_{8}(B^{+}\eta^{\prime},\pi^{+}) +a_{8a}(B^{+}\eta^{\prime},\pi^{+})
       \bigl] R_{\pi}+
         a_{10}(B^{+}\eta^{\prime},\pi^{+}) +a_{10a}(B^{+}\eta^{\prime},\pi^{+})
       \Bigr]-
       \Bigl[ a_{4}^{c}(B^{+}\eta^{\prime},\pi^{+})
           \nonumber \\
    && +\bigl[ a_{6}^{c}(B^{+}\eta^{\prime},\pi^{+})
       +a_{8}(B^{+}\eta^{\prime},\pi^{+}) +a_{8a}(B^{+}\eta^{\prime},\pi^{+})
        \bigr]
      R_{\pi}
    \biggl. \Bigl. +
         a_{10}(B^{+}\eta^{\prime},\pi^{+}) +a_{10a}(B^{+}\eta^{\prime},\pi^{+})
        \Bigr]
       \biggr\};
\end{eqnarray}
\end{flushleft}
\begin{flushleft}
\begin{eqnarray}
{\cal M}(B^{0}\to \eta^{(\prime)}\pi^{0})
  &=&-i \frac{G_{F}}{2} M_{B}^{2} F^{B\to\pi}(m^{2}_{\eta^{(\prime)}}) \lambda V_{cb}
    \biggl\{
         R_{u} e^{i\gamma}
       \Bigl[ a_{2}(B^{0}\pi^{0},\eta^{(\prime)}) f^{u}_{\eta^{(\prime)}}+
         \bigl[ a_{3}(B^{0}\pi^{0},\eta^{(\prime)})-a_{5}(B^{0}\pi^{0},\eta^{(\prime)})
          \bigr]
          (2f^{u}_{\eta^{(\prime)}}
       \Bigr.
     \biggr. \nonumber \\
  && +f^{s}_{\eta^{(\prime)}})+a_{4}^{u}(B^{0}\pi^{0},\eta^{(\prime)})
    f^{u}_{\eta^{(\prime)}}+
     \bigl[ a_{6}^{u}(B^{0}\pi^{0},\eta^{(\prime)})
        -\frac{1}{2}a_{8}(B^{0}\pi^{0},\eta^{(\prime)})+a_{8a}(B^{0}\pi^{0},\eta^{(\prime)})
      \bigr]
         R^{d}_{\eta^{(\prime)}} f^{u}_{\eta^{(\prime)}}
        \nonumber \\
  &&\Bigl. \bigl. +
         \frac{1}{2}\bigl[ a_{9}(B^{0}\pi^{0},\eta^{(\prime)})
         -a_{7}(B^{0}\pi^{0},\eta^{(\prime)}) \bigr]
     (f^{u}_{\eta^{(\prime)}}-f^{s}_{\eta^{(\prime)}})
    +\bigl[ -\frac{1}{2}a_{10}(B^{0}\pi^{0},\eta^{(\prime)})
       +a_{10a}(B^{0}\pi^{0},\eta^{(\prime)})
     \bigr] f^{u}_{\eta^{(\prime)}}
     \Bigr] -
           \nonumber \\
  && \Bigl[ \bigl[ a_{3}(B^{0}\pi^{0},\eta^{(\prime)})
                -a_{5}(B^{0}\pi^{0},\eta^{(\prime)})
                 \bigr](2f^{u}_{\eta^{(\prime)}}+f^{s}_{\eta^{(\prime)}})
       +a_{4}^{c}(B^{0}\pi^{0},\eta^{(\prime)})  f^{u}_{\eta^{(\prime)}}
\nonumber\\
  && +
         \left[ a_{6}^{c}(B^{0}\pi^{0},\eta^{(\prime)})
           -\frac{1}{2}a_{8}(B^{0}\pi^{0},\eta^{(\prime)})
          +a_{8a}(B^{0}\pi^{0},\eta^{(\prime)})
         \right] R^{u}_{\eta^{(\prime)}} f^{u}_{\eta^{(\prime)}}
       \nonumber \\
  && +\frac{1}{2}
     \bigl[ a_{9}(B^{0}\pi^{0},\eta^{(\prime)})
       -a_{7}(B^{0}\pi^{0},\eta^{(\prime)})
     \bigr] (f^{u}_{\eta^{(\prime)}}-f^{s}_{\eta^{(\prime)}})
    +
      \bigl[ -\frac{1}{2}a_{10}(B^{0}\pi^{0},\eta^{(\prime)})
         +a_{10a}(B^{0}\pi^{0},\eta^{(\prime)})
      \bigr]
        f^{u}_{\eta^{(\prime)}}
       \Bigr]
      \biggr\} \nonumber \\
   &&+i \frac{G_{F}}{2} M_{B}^{2} F^{B\to\eta^{(\prime)}} (0) f_{\pi} \lambda V_{cb}
      \biggl\{
         R_{u} e^{i\gamma}
       \Bigl[ a_{2}(B^{0}\eta^{(\prime)},\pi^{0})-a_{4}^{u}(B^{0}\eta^{(\prime)},\pi^{0}) -
          \bigl[a_{6}^{u}(B^{0}\eta^{(\prime)},\pi^{0})
             -\frac{1}{2}a_{8}(B^{0}\eta^{(\prime)},\pi^{0})
       \Bigr. \nonumber \\
   && \Bigl. +a_{8a}(B^{0}\eta^{(\prime)},\pi^{0})
              \bigr] R_{\pi} +
        \frac{3}{2} \bigl[ a_{9}(B^{0}\eta^{(\prime)},\pi^{0})
          -a_{7}(B^{0}\eta^{(\prime)},\pi^{0})]
        +\frac{1}{2}a_{10}(B^{0}\eta^{(\prime)},\pi^{0})-a_{10a}(B^{0}\eta^{(\prime)},\pi^{0})
      \Bigr]  \nonumber \\
   && -
         \Bigl[ -a_{4}^{c}(B^{0}\eta^{(\prime)},\pi^{0}) -
          \bigl[ a_{6}^{c}(B^{0}\eta^{(\prime)},\pi^{0})
                -\frac{1}{2}a_{8}(B^{0}\eta^{(\prime)},\pi^{0})
             +a_{8a}(B^{0}\eta^{(\prime)},\pi^{0})
          \bigr] R_{\pi}+
         \frac{3}{2}\bigl[ a_{9}(B^{0}\eta^{(\prime)},\pi^{0})
         \bigr. \nonumber \\
    && \biggl. \Bigr. \bigl.
           -a_{7}(B^{0}\eta^{(\prime)},\pi^{0})
                      \bigr]
           +\frac{1}{2}a_{10}(B^{0}\eta^{(\prime)},\pi^{0})-a_{10a}(B^{0}\eta^{(\prime)},\pi^{0})
        \Bigr]
         \biggr\}; \\
{\cal M}(B^{+}\to \eta'K^{\ast+})&=&{\cal M}(B^{+}\to \eta K^{\ast+})\,\,
(\eta\to\eta' );\\
{\cal M}(B^{0}\to \eta' K^{\ast0})&=&{\cal M}(B^{+}\to \eta K^{\ast0})\,\,
(\eta\to\eta' );\\
{\cal M}(B^{+}\to\rho^{+}\eta^{(\prime)})
   &=& \frac{G_{F}}{2} M_{B}^{2} A_{0}^{B\to\rho} (m^{2}_{\eta^{(\prime)}}) \lambda V_{cb}
     \biggl\{
             R_{u} e^{i\gamma}
       \Bigl[ a_{2}(B^{+}\rho^{+},\eta^{(\prime)}) f^{u}_{\eta^{(\prime)}}+
         \bigl[ a_{3}(B^{+}\rho^{+},\eta^{(\prime)})
            -a_{5}(B^{+}\rho^{+},\eta^{(\prime)})
         \bigr]
       (2f^{u}_{\eta^{(\prime)}}+f^{s}_{\eta^{(\prime)}})
     \Bigr.
     \biggr. \nonumber \\
   &&+a_{4}^{u}(B^{+}\rho^{+},\eta^{(\prime)})f^{u}_{\eta^{(\prime)}}
     -\bigl[ a_{6}^{u}(B^{+}\rho^{+},\eta^{(\prime)})
      -\frac{1}{2}a_{8}(B^{+}\rho^{+},\eta^{(\prime)})
       +a_{8a}(B^{+}\rho^{+},\eta^{(\prime)})
       \bigr] R^{d}_{\eta^{(\prime)}} f^{u}_{\eta^{(\prime)}}
      \nonumber \\
   && +
         \frac{1}{2}
       \bigl[ a_{9}(B^{+}\rho^{+},\eta^{(\prime)})
         -a_{7}(B^{+}\rho^{+},\eta^{(\prime)})
       \bigr] (f^{u}_{\eta^{(\prime)}}-f^{s}_{\eta^{(\prime)}})
        \Bigl.
     +\bigl[-\frac{1}{2}a_{10}(B^{+}\rho^{+},\eta^{(\prime)})
       +a_{10a}(B^{+}\rho^{+},\eta^{(\prime)})
      \bigr]
          f^{u}_{\eta^{(\prime)}}
        \Bigr] \nonumber \\
   &&-
        \Bigl[  \bigl[ a_{3}(B^{+}\rho^{+},\eta^{(\prime)})
         -a_{5}(B^{+}\rho^{+},\eta^{(\prime)})
        \bigr] (2f^{u}_{\eta^{(\prime)}}+f^{s}_{\eta^{(\prime)}})+
          a_{4}^{c}(B^{+}\rho^{+},\eta^{(\prime)})
            f^{u}_{\eta^{(\prime)}}
            -\bigl[ a_{6}^{c}(B^{+}\rho^{+},\eta^{(\prime)})
         \Bigr.
             \nonumber \\
   && -\frac{1}{2}a_{8}(B^{+}\rho^{+},\eta^{(\prime)})
     +a_{8a}(B^{+}\rho^{+},\eta^{(\prime)})
        \bigr]
       R^{d}_{\eta^{(\prime)}} f^{u}_{\eta^{(\prime)}}+
        \frac{1}{2}\bigl[ a_{9}(B^{+}\rho^{+},\eta^{(\prime)})
         -a_{7}(B^{+}\rho^{+},\eta^{(\prime)})
       \bigr] (f^{u}_{\eta^{(\prime)}}-f^{s}_{\eta^{(\prime)}})
        \nonumber \\
    &&\biggl. \Bigl.
       +\bigl[ -\frac{1}{2}a_{10}(B^{+}\rho^{+},\eta^{(\prime)})
        +a_{10a}(B^{+}\rho^{+},\eta^{(\prime)})
       \bigr] f^{u}_{\eta^{(\prime)}}
      \Bigr]
      \bigg\} \nonumber \\
     &&+\frac{G_{F}}{2} M_{B}^{2} f_{\rho} F^{B\to\eta^{(\prime)}} (0) \lambda V_{cb}
        \biggl\{
         R_{u} e^{i\gamma} \Bigl[ a_{1}(B^{+}\eta^{(\prime)},\rho^{+}) +
         a_{4}^{u}(B^{+}\eta^{(\prime)},\rho^{+}) +a_{10}(B^{+}\eta^{(\prime)},\rho^{+})
                           \Bigr.
        \biggr. \nonumber \\
     && \biggl.
       \Bigl. +a_{10a}(B^{+}\eta^{(\prime)},\rho^{+})
        \Bigr]
        - \Bigl[ a_{4}^{u}(B^{+}\eta^{(\prime)},\rho^{+})
        +a_{10}(B^{+}\eta^{(\prime)},\rho^{+})
        +a_{10a}(B^{+}\eta^{(\prime)},\rho^{+})
        \Bigr]
        \biggr\};\\
{\cal M}(B^{0}\to\rho^{0}\eta^{(\prime)})
   &=&-\frac{G_{F}}{2} M_{B}^{2} A_{0}^{B\to\rho} (m^{2}_{\eta^{(\prime)}}) \lambda V_{cb}
    \biggl\{
           R_{u} e^{i\gamma}
      \Bigl[ a_{2}(B^{0}\rho^{0},\eta^{(\prime)}) f^{u}_{\eta^{(\prime)}}+
        \bigl[ a_{3}(B^{0}\rho^{0},\eta^{(\prime)})
         -a_{5}(B^{0}\rho^{0},\eta^{(\prime)})
        \bigr]
        (2f^{u}_{\eta^{(\prime)}}+f^{s}_{\eta^{(\prime)}})
      \Bigr.
     \biggr. \nonumber \\
   &&+
     a_{4}^{u}(B^{0}\rho^{0},\eta^{(\prime)}) f^{u}_{\eta^{(\prime)}}
      -\bigl[ a_{6}^{u}(B^{0}\rho^{0},\eta^{(\prime)})
     -\frac{1}{2}a_{8}(B^{0}\rho^{0},\eta^{(\prime)})
     +a_{8a}(B^{0}\rho^{0},\eta^{(\prime)})
      \bigr] R^{d}_{\eta^{(\prime)}} f^{u}_{\eta^{(\prime)}}
     \nonumber \\
    && \Bigl.+
         \frac{1}{2}
         \bigl[a_{9}(B^{0}\rho^{0},\eta^{(\prime)})
          -a_{7}(B^{0}\rho^{0},\eta^{(\prime)})
         \bigr] (f^{u}_{\eta^{(\prime)}}-f^{s}_{\eta^{(\prime)}})
        +\bigl[ -\frac{1}{2}a_{10}(B^{0}\rho^{0},\eta^{(\prime)})
        +a_{10a}(B^{0}\rho^{0},\eta^{(\prime)})
         \bigr] f^{u}_{\eta^{(\prime)}}
        \Bigr] \nonumber \\
   && -\bigl[ (a_{3}(B^{0}\rho^{0},\eta^{(\prime)})
       -a_{5}(B^{0}\rho^{0},\eta^{(\prime)})\bigr]
      (2f^{u}_{\eta^{(\prime)}}+f^{s}_{\eta^{(\prime)}})+
         a_{4}^{c}(B^{0}\rho^{0},\eta^{(\prime)}) f^{u}_{\eta^{(\prime)}}
      -\bigl[ a_{6}^{c}(B^{0}\rho^{0},\eta^{(\prime)})
      \nonumber \\
   &&-\frac{1}{2}a_{8}(B^{0}\rho^{0},\eta^{(\prime)})
      +a_{8a}(B^{0}\rho^{0},\eta^{(\prime)})
    \bigr] R^{d}_{\eta^{(\prime)}} f^{u}_{\eta^{(\prime)}}+
         \frac{1}{2} \bigl[ a_{9}(B^{0}\rho^{0},\eta^{(\prime)})
          -a_{7}(B^{0}\rho^{0},\eta^{(\prime)})
            \bigr] (f^{u}_{\eta^{(\prime)}}-f^{s}_{\eta^{(\prime)}})
     \nonumber \\
   && \biggl. \Bigl.
      +\bigl[ -\frac{1}{2}a_{10}(B^{0}\rho^{0},\eta^{(\prime)})
     +a_{10a}(B^{0}\rho^{0},\eta^{(\prime)})
      \bigr] f^{u}_{\eta^{(\prime)}}
      \Bigr]
     \biggr\} \nonumber \\
   &&+ \frac{G_{F}}{2} M_{B}^{2} f_{\rho} F^{B\to\eta^{(\prime)}} (0) \lambda V_{cb}
       \biggl\{
                 R_{u} e^{i\gamma}
            \Bigl[ a_{2}(B^{0}\eta^{(\prime)},\rho^{0})-
         a_{4}^{u}(B^{0}\eta^{(\prime)},\rho^{0})
          + \frac{3}{2}
        \bigl[ a_{9}(B^{0}\eta^{(\prime)},\rho^{0})
        +a_{7}(B^{0}\eta^{(\prime)},\rho^{0})
        \bigr]
           \Bigr.
       \biggr. \nonumber \\
     &&\Bigl.  +\frac{1}{2}a_{10}(B^{0}\eta^{(\prime)},\rho^{0})
          -a_{10a}(B^{0}\eta^{(\prime)},\rho^{0})
        \Bigr]-
         \Bigl[ -a_{4}^{c}(B^{0}\eta^{(\prime)},\rho^{0})
          + \frac{3}{2}[a_{9}(B^{0}\eta^{(\prime)},\rho^{0})
             +a_{7}(B^{0}\eta^{(\prime)},\rho^{0})
          \bigr]
         \Bigr. \nonumber \\
      && \biggl. \Bigl.
             +\frac{1}{2}a_{10}(B^{0}\eta^{(\prime)},\rho^{0})
          -a_{10a}(B^{0}\eta^{(\prime)},\rho^{0})
         \Bigr]
       \biggr\};\\
{\cal M}(B^{+}\to\eta K^{+})&=&{\cal M}(B^{+}\to\eta' K^{+})\,\,
(\eta'\to\eta);\\
{\cal M}(B^{0}\to\eta K^{0})&=&{\cal M}(B^{+}\to\eta' K^{0})\,\,
(\eta' \to\eta).
\end{eqnarray}
\end{flushleft}
\end{scriptsize}

where $R_u =\frac{1-\lambda^{2}/2}{\lambda}|\frac{V_{ub}}{V_{cb}}|$,
$R^{d}_{\eta^{(\prime)}} =\frac{M^2_{\eta^{(\prime)}} }{m_s m_b }
(1-\frac{f^d_{\eta^{(\prime)}} }{f^s_{\eta^{(\prime)}}})$,
and
the corrections to the decays from SHSM can be read from
equations in Sec.3.2.

\begin{center}
{\bf Appendix B }
\end{center}
The integrals  $F^{\eta^{(\prime)}}_{tw2}$, $F^{\eta^{(\prime)}}_{tw3}$
and $F^{\eta^{(\prime)}}_{BPV1}$ are given by
\begin{eqnarray}
F^{\eta^{(\prime)}}_{tw2}&=&12\int^{1}_{0}
dz \frac{\phi_{B}(z)}{z} Y_{1}( z,k_{\eta^{(\prime)}}, m_{\eta^{(\prime)}}),
 \\
F^{\eta^{(\prime)}}_{tw3} &=&2\int^{1}_{0}
dz \frac{\phi_{B}(z)}{z (1-z)} Y_{2}( z,k_{\eta^{(\prime)}}, m_{\eta^{(\prime)}}),
\\
F^{\eta^{(\prime)}}_{BPV1}&=& F^{\eta^{(\prime)}}_{tw2}
\end{eqnarray}
and  $k_{\eta^{(\prime)}} = m_{B}^{2}/m^2_{\eta^{(\prime)} }$.
The functions
$Y_{1}(z, k, m )$ and $Y_{2} (z, k, m )$  are given by
\begin{eqnarray}
Y_{1}(z, k, m )=& &\frac{1}{{\left( k-1 \right) }^2\,m^2}
\Biggl[
      -2\,k\,\left( 1 - z \right)  -
     k^2\,{\left( 1 - z \right) }^2
       + 2\,\left( 1 - k\,z \right)
\Biggr. \nonumber \\
 &&+
      3\,k\,\left( 1 - z \right) \,\left( 1 - k\,z \right)  -
     2\,{\left( 1 - k\,z \right) }^2 -
     2\,\ln (1 - k\,\left( 1 - z \right) ) \nonumber\\
&&+
  3\,k\,\left( 1 - z \right) \,\ln (1 - k\,\left( 1 - z \right) ) -
     k^3\,{\left( 1 - z \right) }^3\,\ln (1 - k\,\left( 1 - z \right) )
     \nonumber \\
&&+
     k^3\,{\left( 1 - z \right) }^3\,\ln (-k\,\left( 1 - z \right) ) +
     2\,\ln (k\,z) - 3\,k\,\left( 1 - z \right) \,\ln (k\,z) \nonumber \\
&&+
     3\,k\,\left( 1 - z \right) \,{\left( 1 - k\,z \right) }^2\,\ln (k\,z) -
     2\,{\left( 1 - k\,z \right) }^3\,\ln (k\,z)\nonumber \\
\Biggl.  &&-
     3\,k\,\left( 1 - z \right) \,{\left( 1 - k\,z \right) }^2\,\ln (-1 + k\,z) +
     2\,{\left( 1 - k\,z \right) }^3\,\ln (-1 + k\,z)
\biggr], \\
Y_{2}(z, k, m)=&& \frac{1}{\left(k-1 \right) \,m^2}
    \Biggl[
        3 \biggl(    -1 + k\,\left( 1 - z \right)  + k\,z
        \biggr.
            \Biggr. \nonumber \\
   && - \ln (1 - k\,\left( 1 - z \right) ) -
      2\,\left( 1 - z \right) \,\ln (1 - k\,\left( 1 - z \right) )
        \biggr.
     \nonumber \\
   &&+
      2\,k\,\left( 1 - z \right) \,\ln (1 - k\,\left( 1 - z \right) ) +
      2\,k\,{\left( 1 - z \right) }^2\,\ln (1 - k\,\left( 1 - z \right) )
   \nonumber \\
    &&- k^2\,{\left( 1 - z \right) }^2\,\ln (1 - k\,\left( 1 - z \right) )
     - 2\,k\,{\left( 1 - z \right) }^2\,\ln (-k\,\left( 1 - z \right) )
  \nonumber \\
 &&+ k^2\,{\left( 1 - z \right) }^2\,\ln (-k\,\left( 1 - z \right) ) +
      2\,\left( 1 - z \right) \,\ln (1 - k\,\left( 1 - z \right)  - z)
\nonumber \\
   &&-
     2\,k\,\left( 1 - z \right) \,
       \ln (1 - k\,\left( 1 - z \right)  - z) \nonumber \\
   &&-
      2\,{\left( 1 - z \right) }^2\,\ln (-k\,\left( 1 - z \right) )\,
       \ln (\frac{1 - k\,\left( 1 - z \right)  - z}{1 - z}) \nonumber \\
   && +
      2\,k\,{\left( 1 - z \right) }^2\,\ln (-k\,\left( 1 - z \right) )\,
       \ln (\frac{1 - k\,\left( 1 - z \right)  - z}{1 - z})  \nonumber \\
   &&+
      2\,{\left( 1 - z \right) }^2\,\ln (1 - k\,\left( 1 - z \right) )\,
       \ln (\frac{-\left( 1 - k\,\left( 1 - z \right)  - z \right) }{z})
   \nonumber \\
   &&-
      2\,k\,{\left( 1 - z \right) }^2\,\ln (1 - k\,\left( 1 - z \right) )\,
       \ln (\frac{-\left( 1 - k\,\left( 1 - z \right)  - z \right) }{z}) +
      \ln (k\,z) \nonumber \\
   &&+ 2\,\left( 1 - z \right) \,\ln (k\,z) -
      2\,k\,\left( 1 - z \right) \,\ln (k\,z) -
      2\,\left( 1 - z \right) \,\left( 1 - k\,z \right) \,\ln (k\,z)
       \nonumber \\
   &&+
      2\,k\,\left( 1 - z \right) \,\left( 1 - k\,z \right) \,\ln (k\,z) -
      {\left( 1 - k\,z \right) }^2\,\ln (k\,z)
    \nonumber \\
  &&+
      2\,\left( 1 - z \right) \,\left( 1 - k\,z \right) \,\ln (-1 + k\,z) -
      2\,k\,\left( 1 - z \right) \,\left( 1 - k\,z \right) \,\ln (-1 + k\,z)
    \nonumber \\
   &&+
      {\left( 1 - k\,z \right) }^2\,\ln (-1 + k\,z) -
      2\,\left( 1 - z \right) \,\ln (-z + k\,z)
      \nonumber \\
   &&+
      2\,k\,\left( 1 - z \right) \,\ln (-z + k\,z) +
      2\,{\left( 1 - z \right) }^2\,\ln (-1 + k\,z)\,
       \ln (\frac{-z + k\,z}{1 - z}) \nonumber \\
   &&-
      2\,k\,{\left( 1 - z \right) }^2\,\ln (-1 + k\,z)\,
       \ln (\frac{-z + k\,z}{1 - z}) -
      2\,{\left( 1 - z \right) }^2\,\ln (k\,z)\,
       \ln (\frac{-\left( -z + k\,z \right) }{z})   \nonumber \\
   &&+
      2\,k\,{\left( 1 - z \right) }^2\,\ln (k\,z)\,
       \ln (\frac{-\left( -z + k\,z \right) }{z})
    -
      4\,{\left( 1 - z \right) }^2\, Li_{2}(k) \nonumber \\
  &&+
      4\,k\,{\left( 1 - z \right) }^2\,Li_{2}(k) +
      2\,{\left( 1 - z \right) }^2\,
       Li_{2}\left( \frac{-\left( -1 + k\,\left( 1 - z \right)  \right) }
         {z})\right)
       \nonumber \\
 &&-   2\,k\,{\left( 1 - z \right) }^2\,
       Li_{2}(\frac{-\left( -1 + k\,\left( 1 - z \right)  \right) }
         {z}) + 2\,{\left( 1 - z \right) }^2\,
       Li_{2}\left( \frac{1 - k\,z}{1 - z}\right) \nonumber \\
 && \Biggl. \biggl.-
      2\,k\,{\left( 1 - z \right) }^2\,
       Li_{2}\left(\frac{1 - k\,z}{1 - z}\right)
       \biggr)
    \Biggr].
\end{eqnarray}
Where $Li_{2}(x)$ is the dilogarithm or Spence function.

\newpage

\begin{center} {\bf Figure captions} \end{center}

FIG.1 Order $\alpha_s$  non-factorizable contributions
in $B\to M_1 M_2$ decays.

FIG.2 The triangle Feynman diagrams for the transition form-factors
of $g^{\ast}g^{\ast}-\eta'$ and $g^{\ast}g^{\ast}-\eta$.

FIG.3 The Feynman diagrams of the spectator hard scattering mechanism
for B decays to $\eta' M$, where M is a light pseudoscalar or vector meson.

FIG.4 Branching ratios for B decays  are shown as
 curves as a function of $\gamma$ in units of $10^{-6}$. The dashing curves
are the results of conventional mechanisms, the solid curves are our estimations
with incorporating SHSM contributions.
The branching ratios  measured by CLEO  Collaboration are shown by horizontal
solid lines. The thicker solid horizontal
lines are  its center values, thin horizonal lines are its error bars.

FIG.5 Direct CP violations in B decays involving $\eta^{(\prime)}$.
Dashing curves are the predictions of conventional mechanism estimated with
the QCD improved factorization approach.  Solid curves are results
when SHSM contributions added.

\newpage

\begin{figure}[htbp]
 \scalebox{0.7}{
 {
   \begin{picture}(140,120)(-30,0)
    \ArrowLine(0,40)(30,40)
    \ArrowLine(30,40)(60,40)
    \ArrowLine(60,40)(90,40)
    \ArrowLine(90,20)(0,20)
    \Gluon(30,40)(46,87){4}{4} \Vertex(30,40){1.5} \Vertex(46,87){1.5}
    \Line(58,42)(62,38)
    \Line(58,38)(62,42)
    \ArrowLine(40,105)(60,45)
    \ArrowLine(60,45)(80,105)
    \put(-20,28){$\bar{B}$}
    \put(90,28){$M_{1}$}
    \put(58,110){$M_{2}$}
    \put(55,28){\small{${\cal O}_i $}}
    \put(0,45){\small{$b$}}
    \put(45,0){(a)}
 \end{picture}
 }}
 \scalebox{0.7}{
 {
   \begin{picture}(140,120)(-30,0)
    \ArrowLine(0,40)(30,40)
    \ArrowLine(30,40)(60,40)
    \ArrowLine(60,40)(90,40)
    \ArrowLine(90,20)(0,20)
    \Gluon(30,40)(74,87){4}{6} \Vertex(30,40){1.5} \Vertex(74,87){1.5}
    \Line(58,42)(62,38)
    \Line(58,38)(62,42)
    \ArrowLine(40,105)(60,45)
    \ArrowLine(60,45)(80,105)
    \put(-20,28){$\bar{B}$}
    \put(90,28){$M_{1}$}
    \put(58,110){$M_{2}$}
    \put(55,28){\small{${\cal O}_i $}}
    \put(0,45){\small{$b$}}
    \put(45,0){(b)}
 \end{picture}
 }}
 \scalebox{0.7}{
 {
   \begin{picture}(140,120)(-30,0)
    \ArrowLine(0,40)(20,40)
    \ArrowLine(20,40)(60,40)
    \ArrowLine(60,40)(90,40)
    \ArrowLine(90,20)(0,20)
    \Gluon(54,87)(65,40){3}{6} \Vertex(65,40){1.5} \Vertex(54,87){1.5}
    \Line(38,42)(42,38)
    \Line(38,38)(42,42)
    \ArrowLine(20,105)(40,45)
    \ArrowLine(40,45)(60,105)
    \put(-20,28){$\bar{B}$}
    \put(90,28){$M_{1}$}
    \put(58,110){$M_{2}$}
    \put(35,28){\small{${\cal O}_i $}}
    \put(0,45){\small{$b$}}
    \put(45,0){(c)}
 \end{picture}
 }}
 \scalebox{0.7}{
 {
   \begin{picture}(140,120)(-30,0)
    \ArrowLine(0,40)(20,40)
    \ArrowLine(20,40)(60,40)
    \ArrowLine(60,40)(90,40)
    \ArrowLine(90,20)(0,20)
    \Gluon(26,87)(65,40){3}{6} \Vertex(65,40){1.5} \Vertex(54,87){1.5}
    \Line(38,42)(42,38)
    \Line(38,38)(42,42)
    \ArrowLine(20,105)(40,45)
    \ArrowLine(40,45)(60,105)
    \put(-20,28){$\bar{B}$}
    \put(90,28){$M_{1}$}
    \put(58,110){$M_{2}$}
    \put(35,28){\small{${\cal O}_i $}}
    \put(0,45){\small{$b$}}
    \put(45,0){(d)}
 \end{picture}
 }}

\vspace{1cm}
\scalebox{0.7}{
 {
   \begin{picture}(140,120)(-30,0)
    \ArrowLine(90,20)(-5,20)
    \ArrowLine(-5,50)(20,50)
    \ArrowLine(20,50)(30,100)
    \ArrowLine(70,50)(90,50)
    \ArrowLine(80,100)(70,50)
   \put(35,50){\circle{27}}
    \Gluon(50,50)(70,50){3}{3} \Vertex(50,50){1.5} \Vertex(70,50){1.5}
    \Line(18,52)(22,48)
    \Line(18,48)(22,52)
    \put(-20,28){$\bar{B}$}
    \put(90,28){$M_{1}$}
    \put(58,110){$M_{2}$}
    \put(23, 45){\small{${\cal O}_i $}}
    \put(0,54){\small{$b$}}
    \put(45,0){(e)}
 \end{picture}
 }}
\scalebox{0.7}{
 {
   \begin{picture}(140,120)(-30,0)
    \ArrowLine(90,20)(-5,20)
    \ArrowLine(-5,50)(20,50)
    \ArrowLine(20,50)(30,100)
    \ArrowLine(70,50)(90,50)
    \ArrowLine(80,100)(70,50)
    \Gluon(20,50)(70,50){3}{6} \Vertex(20,50){1.5} \Vertex(70,50){1.5}
    \Line(18,52)(22,48)
    \Line(18,48)(22,52)
    \put(-20,28){$\bar{B}$}
    \put(90,28){$M_{1}$}
    \put(58,110){$M_{2}$}
    \put(19, 40){\small{${\cal O}_g $}}
    \put(0,54){\small{$b$}}
    \put(45,0){(f)}
 \end{picture}
 }}
\scalebox{0.7}{
 {
   \begin{picture}(140,120)(-30,0)
    \ArrowLine(0,40)(60,40)
    \ArrowLine(60,40)(90,40)
    \ArrowLine(90,20)(0,20)
    \Gluon(30,20)(46,87){4}{10} \Vertex(30,20){1.5} \Vertex(46,87){1.5}
    \Line(58,42)(62,38)
    \Line(58,38)(62,42)
    \ArrowLine(40,105)(60,45)
    \ArrowLine(60,45)(80,105)
    \put(-20,28){$\bar{B}$}
    \put(90,28){$M_{1}$}
    \put(58,110){$M_{2}$}
    \put(55,28){\small{${\cal O}_i $}}
    \put(0,45){\small{$b$}}
    \put(45,0){(g)}
 \end{picture}
 }}
\scalebox{0.7}{
 {
   \begin{picture}(140,120)(-30,0)
    \ArrowLine(0,40)(20,40)
    \ArrowLine(20,40)(90,40)
    \ArrowLine(90,20)(0,20)
    \Gluon(54,87)(65,20){3}{6} \Vertex(65,20){1.5} \Vertex(54,87){1.5}
    \Line(38,42)(42,38)
    \Line(38,38)(42,42)
    \ArrowLine(20,105)(40,45)
    \ArrowLine(40,45)(60,105)
    \put(-20,28){$\bar{B}$}
    \put(90,28){$M_{1}$}
    \put(58,110){$M_{2}$}
    \put(35,28){\small{${\cal O}_i $}}
    \put(0,45){\small{$b$}}
    \put(45,0){(h)}
 \end{picture}
 }}
\caption{ }
\end{figure}

\newpage

\begin{figure}[htbp]
\scalebox{0.9}{
 {
    \begin{picture}(150,100)(-50,0)
    \Gluon(20,20)(60,20){3}{6} \Vertex(60,20){1.5}
     \Gluon(20,80)(60,80){3}{6} \Vertex(60,80){1.5}
    \ArrowLine(60,20)(60,80)
\ArrowLine(60,80)(110,50)\Vertex(110,50){2}
\ArrowLine(110,50)(60,20)
\put(120,50){$\eta^{\prime}$}
\put(20,90){$Q_1$}
\put(20,5){$Q_2$}
 \put(33,85){$\to$}
 \put(33,25){$\to$}
\put(70,-15){(a)}
\put(150,-60){$\bf{Figure~~2}$ }
\end{picture}
 }}
\scalebox{0.9}{
 {
    \begin{picture}(150,100)(-50,0)
    \Gluon(20,20)(60,80){3}{10} \Vertex(60,20){1.5}
     \Gluon(20,80)(60,20){3}{10} \Vertex(60,80){1.5}
    \ArrowLine(60,20)(60,80)
\ArrowLine(60,80)(110,50)\Vertex(110,50){2}
\ArrowLine(110,50)(60,20)
\put(120,50){$\eta^{\prime}$}
\put(20,90){$Q_1$}
\put(20,5){$Q_2$}
\put(70,-15){(b)}
 \end{picture}
 }}
 \end{figure}

\vspace{5cm}
\begin{figure}[htbp]
\scalebox{0.9}{
 {
    \begin{picture}(150,100)(-50,0)
    \ArrowLine(100,20)(20,20)\Vertex(100,20){2}
\put(20,65){$\bar{z}P_B$}
\ArrowLine(160,20)(100,20)
\put(145,65){$\bar{y}P_M$}
\ArrowLine(20,60)(90,60)\Vertex(90,60){2}
\put(83,49){${\cal O}_{g}$ }
\put(20,25){$zP_B$}
\put(40,23){$\to$}
\ArrowLine(90,60)(160,60)
\put(130,23){$\to$}
\put(145,25){$yP_M$}
\Gluon(90,60)(150,120){3}{6}
     \Gluon(100,20)(155,115){3}{9} \Vertex(152.5,117.5){3.5}
    \put(160,125){$\eta^{\prime}$}
\put(10,40){$B$}
\put(165,40){$M$}
\put(70,-15){(a)}
\put(150,-60){$\bf{Figure~~3}$ }
\end{picture}
 }}
\hspace{20mm}
 \scalebox{0.9}{
 {
    \begin{picture}(150,100)(-50,0)
    \ArrowLine(100,20)(20,20)\Vertex(100,20){2}
\put(20,65){$\bar{z}P_B$}
\ArrowLine(160,20)(100,20)
\put(145,65){$\bar{y}P_M$}
\ArrowLine(20,60)(90,60)\Vertex(90,60){2}
\put(83,49){${\cal O}_{i}$ }
\put(20,25){$zP_B$}
\put(45,23){$\to$}
\ArrowLine(90,60)(160,60)
\put(130,23){$\to$}
\put(145,25){$yP_M$}
\Gluon(102,86.5)(150,120){3}{4}\Vertex(102,86.5){1.3}
  \put(90,77){\circle{30}}
   \Gluon(100,20)(155,115){3}{9} \Vertex(152.5,117.5){3.5}
    \put(160,125){$\eta^{\prime}$}
\put(10,40){$B$}
\put(165,40){$M$}
\put(70,-15){(b)}
\end{picture}
 }}
\end{figure}

\setlength{\oddsidemargin}{-10mm}
\begin{figure}[htbp]
\vspace{-30mm}
\scalebox{1.0}{\psfig{file=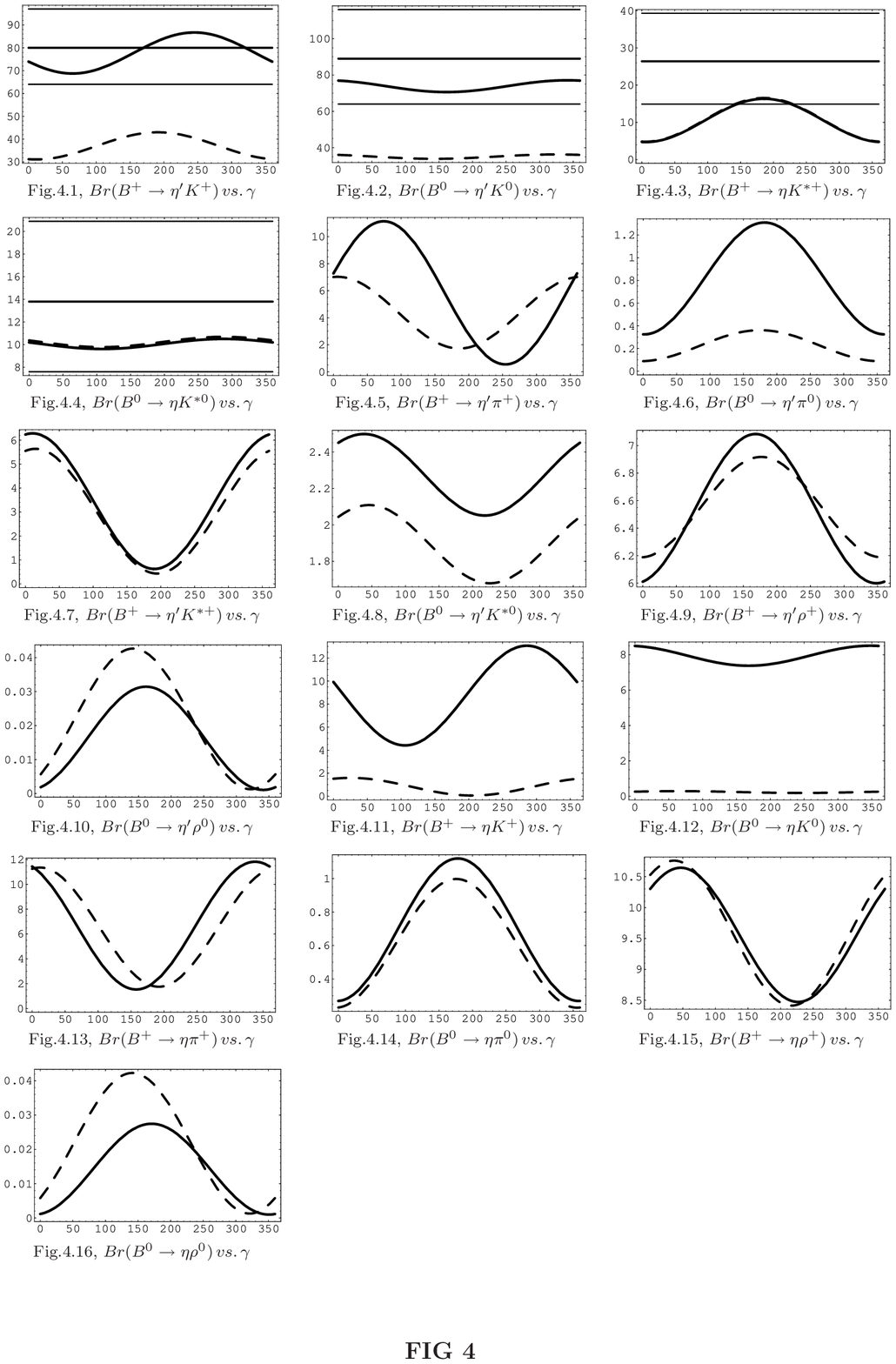}}
 \end{figure}
\newpage
\begin{figure} [htbp]
\scalebox{0.9}{\psfig{file=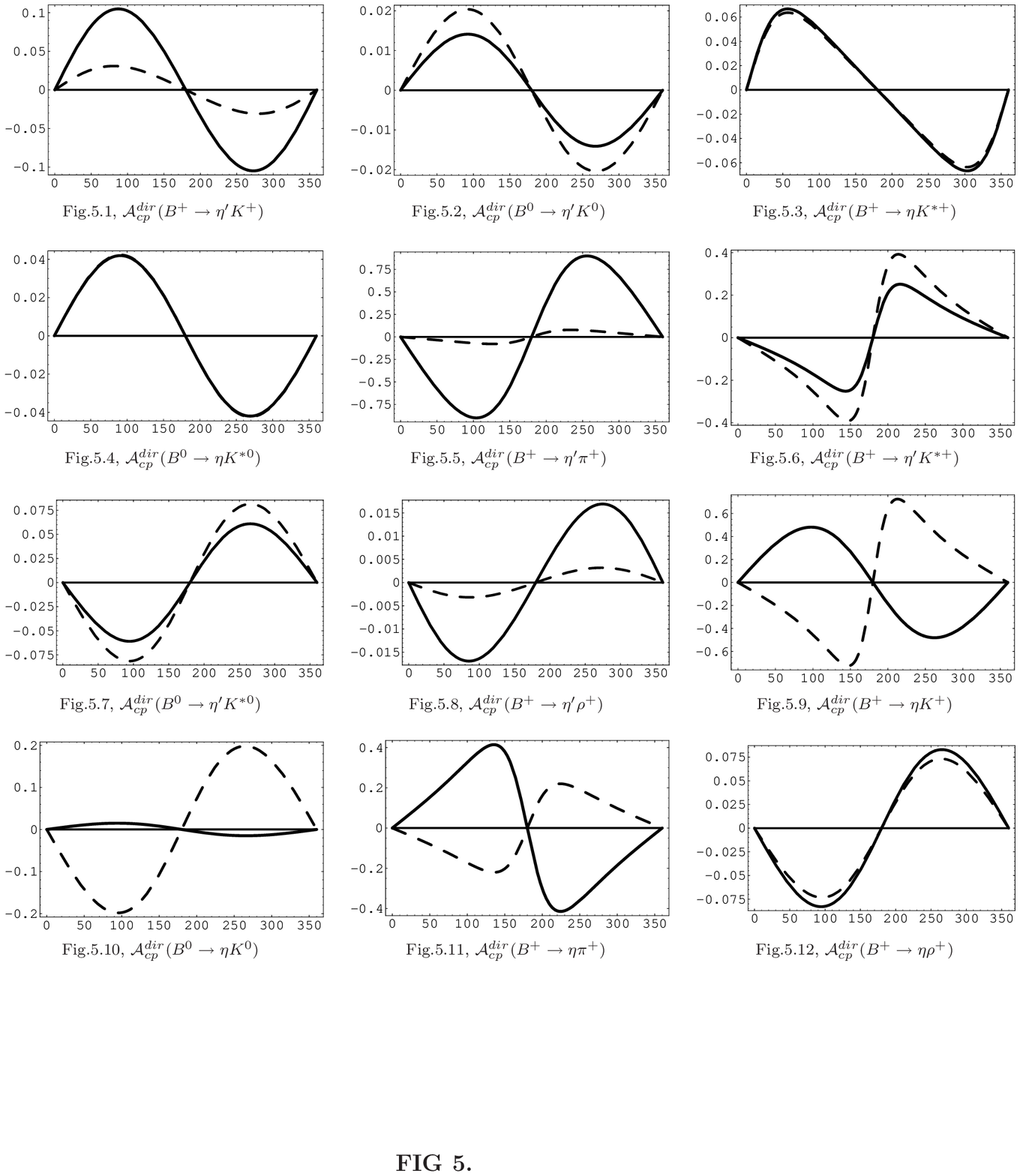}}
 \end{figure}
\end{document}